\documentclass[12pt]{article}
\PassOptionsToPackage{unicode}{hyperref}
\PassOptionsToPackage{hyphens}{url}
\PassOptionsToPackage{dvipsnames,svgnames,x11names}{xcolor}

\usepackage{amsmath,amssymb,amsthm}
\usepackage{iftex}
\ifPDFTeX
  \usepackage[T1]{fontenc}
  \usepackage[utf8]{inputenc}
  \usepackage{textcomp} 
\else 
  \usepackage{unicode-math}
  \defaultfontfeatures{Scale=MatchLowercase}
  \defaultfontfeatures[\rmfamily]{Ligatures=TeX,Scale=1}
\fi
\usepackage{lmodern}
\ifPDFTeX\else  
\fi
\IfFileExists{upquote.sty}{\usepackage{upquote}}{}
\IfFileExists{microtype.sty}{
  \usepackage[]{microtype}
  \UseMicrotypeSet[protrusion]{basicmath} 
}{}
\makeatletter
\@ifundefined{KOMAClassName}{
  \IfFileExists{parskip.sty}{%
    \usepackage{parskip}
  }{
    \setlength{\parindent}{0pt}
    \setlength{\parskip}{6pt plus 2pt minus 1pt}}
}{
  \KOMAoptions{parskip=half}}
\makeatother
\usepackage{xcolor}
\makeatletter
\ifx\paragraph\undefined\else
  \let\oldparagraph\paragraph
  \renewcommand{\paragraph}{
    \@ifstar
      \xxxParagraphStar
      \xxxParagraphNoStar
  }
  \newcommand{\xxxParagraphStar}[1]{\oldparagraph*{#1}\mbox{}}
  \newcommand{\xxxParagraphNoStar}[1]{\oldparagraph{#1}\mbox{}}
\fi
\ifx\subparagraph\undefined\else
  \let\oldsubparagraph\subparagraph
  \renewcommand{\subparagraph}{
    \@ifstar
      \xxxSubParagraphStar
      \xxxSubParagraphNoStar
  }
  \newcommand{\xxxSubParagraphStar}[1]{\oldsubparagraph*{#1}\mbox{}}
  \newcommand{\xxxSubParagraphNoStar}[1]{\oldsubparagraph{#1}\mbox{}}
\fi
\makeatother

\usepackage{longtable,booktabs,array}
\usepackage{calc} 
\usepackage{etoolbox}
\makeatletter
\patchcmd\longtable{\par}{\if@noskipsec\mbox{}\fi\par}{}{}
\makeatother
\IfFileExists{footnotehyper.sty}{\usepackage{footnotehyper}}{\usepackage{footnote}}
\makesavenoteenv{longtable}
\usepackage{graphicx}
\makeatletter
\def\maxwidth{\ifdim\Gin@nat@width>\linewidth\linewidth\else\Gin@nat@width\fi}
\def\maxheight{\ifdim\Gin@nat@height>\textheight\textheight\else\Gin@nat@height\fi}
\makeatother
\setkeys{Gin}{width=\maxwidth,height=\maxheight,keepaspectratio}
\makeatletter
\def\fps@figure{htbp}
\makeatother

\makeatletter
\@ifpackageloaded{caption}{}{\usepackage{caption}}
\AtBeginDocument{%
\ifdefined\contentsname
  \renewcommand*\contentsname{Table of contents}
\else
  \newcommand\contentsname{Table of contents}
\fi
\ifdefined\listfigurename
  \renewcommand*\listfigurename{List of Figures}
\else
  \newcommand\listfigurename{List of Figures}
\fi
\ifdefined\listtablename
  \renewcommand*\listtablename{List of Tables}
\else
  \newcommand\listtablename{List of Tables}
\fi
\ifdefined\figurename
  \renewcommand*\figurename{Figure}
\else
  \newcommand\figurename{Figure}
\fi
\ifdefined\tablename
  \renewcommand*\tablename{Table}
\else
  \newcommand\tablename{Table}
\fi
}
\@ifpackageloaded{float}{}{\usepackage{float}}
\floatstyle{ruled}
\@ifundefined{c@chapter}{\newfloat{codelisting}{h}{lop}}{\newfloat{codelisting}{h}{lop}[chapter]}
\floatname{codelisting}{Listing}

\makeatother
\makeatletter
\@ifpackageloaded{caption}{}{\usepackage{caption}}
\@ifpackageloaded{subcaption}{}{\usepackage{subcaption}}
\makeatother

\ifLuaTeX
  \usepackage{selnolig}  
\fi
\usepackage[]{natbib}
\usepackage{bookmark}

\IfFileExists{xurl.sty}{\usepackage{xurl}}{} 
\hypersetup{
  pdftitle={Sampling-Based Batch Sequential Design by Stein Variational Gradient Descent},
  pdfauthor={},
  pdfkeywords={Computer Experiment; Constrained Sampling; Fully Sequential Design; Space-filling Design},
  colorlinks=true,
  linkcolor={blue},
  filecolor={Maroon},
  citecolor={Blue},
  urlcolor={Blue},
  pdfcreator={LaTeX via pandoc}}
  
\usepackage{mathtools}          
\usepackage{aliascnt}           

\usepackage{bm}                 

\usepackage{subcaption}

\usepackage{algorithm}
\usepackage{algpseudocode}      
\algnewcommand{\Optional}{\item[\textbf{Optional:}]}

\usepackage{setspace}           

\usepackage{comment}

\usepackage[nameinlink]{cleveref}
\crefname{figure}{Figure}{Figures}
\Crefname{figure}{Figure}{Figures}
\crefname{section}{Section}{Sections}
\Crefname{section}{Section}{Sections}
\crefname{subsection}{Section}{Sections}
\Crefname{subsection}{Section}{Sections}
\crefname{subsubsection}{Section}{Sections}
\Crefname{subsubsection}{Section}{Sections}
\crefname{algorithm}{Algorithm}{Algorithms}
\Crefname{algorithm}{Algorithm}{Algorithms}

\usepackage{enumitem}

\newaliascnt{lemma}{theorem}

\aliascntresetthe{lemma}

\newaliascnt{proposition}{theorem}

\aliascntresetthe{proposition}

\newaliascnt{corollary}{theorem}

\aliascntresetthe{corollary}

\theoremstyle{definition}
\newaliascnt{definition}{theorem}

\aliascntresetthe{definition}

\theoremstyle{remark}

\DeclareMathOperator*{\argmax}{arg\,max}
\DeclareMathOperator*{\argmin}{arg\,min}
\DeclareMathOperator{\Cov}{Cov}

\newcommand{\norm}[1]{\left\lVert#1\right\rVert}
\newcommand{\diff}{\mathop{}\!\mathrm{d}}  

\newcommand{\T}{^{\mkern-1mu\mathsf{T}}}  
\newcommand{\GP}{\mathcal{GP}}    

\ExplSyntaxOn
\tl_map_inline:nn { abcdefghijklmnopqrstuvwxyz }
{ \cs_set:cpn { v #1 } { { \bm { #1 } } } }
\tl_map_inline:nn { ABCDEFGHIJKLMNOPQRSTUVWXYZ }
{ \cs_set:cpn { m #1 } { { \bm { #1 } } } }
\tl_map_inline:nn { EIRV }
{\cs_new:cpn { bb#1 } { \mathbb{#1} }}
\tl_map_inline:nn { ADFGHILMNOTXY }
{\cs_new:cpn { cal#1 } { \mathcal{#1} }}
\ExplSyntaxOff

\newcommand{\btheta}{{\bm{\theta}}}

\newcommand{\bphi}{{\bm{\phi}}}

\newcommand{\defeq}{\coloneqq}

\newcommand{\given}{\mathrel{}\middle\vert\mathrel{}}

\allowdisplaybreaks

\newif\ifdraft
\draftfalse

\ifdraft

\else

\fi
\usepackage{multibib}
\newcites{supp}{Supplementary Materials References} 

\newcommand{\anon}{1}

\begin{document}

\def\spacingset#1{\renewcommand{\baselinestretch}%
{#1}\small\normalsize} \spacingset{1}


\if1\anon
{
  \title{\bf Sampling-Based Batch Sequential Design by Stein Variational Gradient Descent\textsuperscript{\textdaggerdbl}}
  \author{Penghui Fu\textsuperscript{\textdagger}\\
    School of Data Science, The Chinese University of Hong Kong, Shenzhen\\
    Xiaoxian Ding\textsuperscript{\textdagger} \\
    School of Data Science, The Chinese University of Hong Kong, Shenzhen\\
    Chunlin Ji \\
    Kuang-Chi Institute of Advanced Technology \\
    Jianhua Z. Huang\\
    School of Data Science, The Chinese University of Hong Kong, Shenzhen\\
    and \\
    C. F. Jeff Wu\textsuperscript{*}\\
    School of Data Science, The Chinese University of Hong Kong, Shenzhen
    } 
  \maketitle
  \textsuperscript{\textdagger}Joint first authors.
  \textsuperscript{*}Corresponding author: \href{mailto:jeffwu@cuhk.edu.cn}{jeffwu@cuhk.edu.cn} \\
  \textsuperscript{\textdaggerdbl}This work was supported by Shenzhen Kuang-Chi Cutting-Edge Technology Co., Ltd. through a research project and by the National Natural Science Foundation of China (NSFC) under Grant No.~W2631001.
} \fi

\if0\anon
{
  \bigskip
  \bigskip
  \bigskip
  \begin{center}
    {\LARGE\bf Sampling-Based Batch Sequential Design by Stein Variational Gradient Descent}
\end{center}
  \medskip
} \fi

\bigskip
\begin{abstract}
Many real-world experimental design problems require a batch of experimental runs across stages, in which multiple points are selected and evaluated at each stage. However, most work in the design literature is focused on fully sequential (point-by-point) methods. This paper proposes a sampling-based framework to systematically convert a fully sequential method to a batch sequential method. In particular, Stein variational gradient descent (SVGD) is adapted to efficiently sample a batch of points from a properly constructed target distribution while balancing the individual utility and the batch diversity. We address challenges that arise in using SVGD for experimental designs, including constrained design regions and near-uniform target distributions. We apply the proposed method to obtain batch versions of the state-of-the-art fully sequential methods, and demonstrate their performance through extensive numerical studies.
\end{abstract}

\noindent%
{\it Keywords:} Computer Experiment; Constrained Sampling; Fully Sequential Design; Space-filling Design.
\vfill

\newpage
\spacingset{1.8} 

\section{Introduction}
\label{sec:intro}
In many experimental situations, we can actively choose which data to collect in stages.
At each stage, data from earlier stages is leveraged to update our understanding of the problem of interest, which, in turn, guides subsequent data collection.
More specifically, a \textit{sequential design} selects the next input point by maximizing its utility given the current data.
In active learning, the utility is called an \textit{acquisition function}, and we use the two terms interchangeably hereafter.
After observing the output, both the model and the acquisition function are updated, and the process repeats.

Sequential designs offer distinct advantages over one-shot optimal and space-filling designs \citep{santner2018design,wu2021experiments}.
Compared to one-shot optimal designs, sequential designs update the model as data are collected, avoid relying on a fixed initial assumption, and are often computationally more efficient.
Compared to space-filling designs, sequential designs do not need to pre-specify the design size and prioritize data collection in most informative regions.
But sequential designs take more time to implement and can be impractical if the investigation is under time pressure.


When multiple experiments can be conducted simultaneously, design points can be selected and evaluated in \textit{batches}, yielding substantial time and cost savings relative to fully sequential (i.e., point-by-point) approaches.
For example, expensive computer experiments can be distributed across multiple CPU cores to reduce wall-clock time \citep{mak2018efficient,chang2021reduced};
deep learning hyperparameter tuning can leverage parallel evaluation to efficiently search high-dimensional parameter spaces \citep{falkner2018bohb};
in industrial quality assurance, testing chambers can evaluate hundreds of parts simultaneously under different stress conditions, thereby substantially amortizing life-testing costs.

While computer experiments literature focuses mostly on fully sequential designs, batch designs have been extensively studied in Bayesian optimization \citep{shah2015parallel,wu2016parallel,crovini2026batch}
and deep active learning \citep{ash2020deep,citovsky2021batch,ren2021survey}. 
While the exact formulations of batch designs depend on tasks and objectives, existing methodologies generally fall into a few templates.
One natural approach is to extend a single-point acquisition function to a batch version.
Such an extension is straightforward by definition for many existing fully sequential methods.
However, this \textit{direct extension} approach often faces significant computational challenges as the batch size increases,
requiring joint optimization over multiple design points and expensive batch acquisition function evaluations.
For a detailed review, see Section~\ref{sec:extension}.

Alternative approaches select batches using single-point utilities.
However, these utilities do not account for the inter-relationships among batch points, so simply selecting the highest-utility points may cause strong correlations and redundancy among points within the batch.
To introduce diversity regularization, existing methods either \textit{explicitly} penalize average utility with a similarity measure \citep{kee2018query,nguyen2024quality}, or \textit{implicitly} encourage diversity using techniques such as determinantal point processes \citep{kathuria2016batched,nava2022diversified} or clustering \citep{ash2020deep,yang2021batch}.

In this article, we propose a sampling-based framework to systematically convert a fully sequential method into a batch sequential method.
By constructing a target density that encodes the single-point acquisition function, we transform the batch design task into a sampling problem.
We then apply and adapt ``Stein variational gradient descent'' \citep{liu2016svgd}, which is abbreviated as SVGD, to sample a batch of points from the target.
SVGD inherently balances utility with diversity and is computationally convenient.
However, SVGD, originally proposed for Bayesian inference, faces \textit{new challenges} when applied to experimental designs.
One challenge is that many design problems involve a constrained region, which makes the vanilla SVGD invalid. 
We borrow ideas from the penalty method in constrained optimization to handle regions with inequality constraints.
Another challenge is that the target distribution may become nearly uniform and therefore non-informative in some cases. 
Consequently, SVGD repeatedly selects similar points, leading to redundancy.
We mitigate this issue with a novel local penalty method, which encourages SVGD to select points away from the existing ones.

The rest of this article is organized as follows.
Section~\ref{sec:background} formulates the batch design problem and reviews batch sequential designs by direct extension.
Section~\ref{sec:approach} proposes the sampling-based framework for batch sequential designs via SVGD, together with penalty-based modifications for constrained design regions and redundancy reduction with existing design points.
Section~\ref{sec:numerical} applies the proposed method to construct batch versions of the active learning method by \citet{mackay1992information} under two surrogate models, presents numerical studies, and compares their performance with fully sequential and batch-by-definition methods.
Section~\ref{sec:conclusion} gives a summary and discusses future work.
All technical derivations are deferred to the Supplement.

\section{Background}\label{sec:background}

\subsection{Problem setup}
Let $\calD_n \defeq \{(\vx_i, y_i)\}_{i=1}^n$ denote the current dataset, where $\vx_i \in \Omega\subset \bbR^d$ is the $i$-th input, $y_i \in \bbR$ is the $i$-th response, and $\Omega$ is the design region. 
For ease of presentation, we focus on the regression setup, although the framework also applies to other problems such as classification.
Denote the true regression function by $f$.
Let $a_n : \Omega \to \bbR$ be an acquisition function which assigns a utility score to each candidate point $\vx$ based on current data $\calD_n$.
A fully sequential method selects the next input point by maximizing the acquisition function, i.e., $\vx^\star = \argmax_{\vx \in \Omega} a_n(\vx)$.
After evaluating $\vx^\star$ and obtaining the response $y^\star$, the dataset is augmented as $\calD_{n+1} = \calD_n \cup \{(\vx^\star, y^\star)\}$, and the procedure repeats.
Our goal is to extend a fully sequential method to a batch sequential method, where at each iteration, $q$ points $\mX_b^\star = (\vx_{n+1}^\star, \dots, \vx_{n+q}^\star)\T$ are selected and evaluated, with a batch size $q\geq 1$.

\subsection{The direct extension approach: a review}\label{sec:extension}

As discussed in Section~\ref{sec:intro}, a natural way to construct a batch design is to extend a single-point acquisition function $a_n(\vx)$ to a batch acquisition $a_n(\mX_b)$ and then select the batch by a joint maximization: $\mX_b^\star = \argmax_{\mX_b} a_n(\mX_b)$.
For many existing fully sequential methods, such extensions are mathematically straightforward but computationally challenging.
For the purpose of referencing, we will refer to this as the \textit{direct extension} approach.
Below, we take commonly used fully sequential criteria in computer experiments for illustration.

\textbf{Variance-based criteria (ALM, ALC):} Active Learning MacKay \citep[ALM,][]{mackay1992information,Seo2000active} selects the point to maximize the posterior variance of the output. 
A natural extension of ALM is to select a batch of points by maximizing the determinant of the posterior covariance matrix, i.e.,
\begin{equation}\label{eq:batch-ALM}
    \argmax_{\vx_{n+1}}\, \bbV\!\left(f(\vx_{n+1})\given\calD_n\right)
    \;\Longrightarrow\;
    \argmax_{\mX_b}\, \det\Cov\!\left(\vf_b\given\calD_n\right),
\end{equation}
where $\vf_b \defeq \bigl(f(\vx_{n+1}), \dots, f(\vx_{n+q})\bigr)\T$ denotes $q$ responses. 
Active Learning Cohn \citep[ALC,][]{cohn1996alc,Seo2000active} targets the maximal reduction of integrated mean-squared prediction error (IMSPE) after the next point is added, which can be naturally extended to the batch setting \citep{zhang2022batch}: 
\begin{equation}\label{eq:batch-ALC}
    \begin{split}
        &\argmax_{\vx_{n+1}}\, \int_{\Omega}\bbV\!\left(f(\vx)\given\calD_n\right)\diff\vx - \int_{\Omega} \bbE_{y_{n+1}}\left[\bbV\!\left(f(\vx)\given\calD_n\cup\{\vx_{n+1},y_{n+1}\}\right)\right]\diff\vx
    \;\\
    \quad \Longrightarrow\;
    &\argmax_{\mX_b}\,\int_{\Omega}\bbV\!\left(f(\vx)\given\calD_n\right)\diff\vx -  \int_{\Omega} \bbE_{\vy_b}\left[\bbV\!\left(f(\vx)\given\calD_n\cup\{\mX_b,\vy_b\}\right)\right] \diff\vx.
    \end{split}
\end{equation}
Here, $y_{n+1}$ and $\vy_b$ denote the corresponding responses at inputs $\vx_{n+1}$ and $\mX_b$, respectively, and $\bbE_{y_{n+1}}$ and $\bbE_{\vy_b}$ are expectations taken with respect to $p(y_{n+1}\mid \vx_{n+1},\calD_n)$ and $p(\vy_b\mid \mX_b,\calD_n)$, respectively.

\textbf{Optimization-driven criteria ($q$-EI):} For black-box minimization problems, one of the most popular criteria is the Expected Improvement criterion \citep[EI,][]{jones1998efficient}, which can be extended to $q$-EI \citep{ginsbourger2010kriging}:
\begin{equation}\label{eq:batch-EI}
     \argmax_{\vx_{n+1}}\,\bbE\left[\left(y^\star_n - f(\vx_{n+1})\right)_+\given\calD_n\right] 
        \Rightarrow\; \argmax_{\mX_b}\,\bbE\left[\left(y^\star_n - \min_{1\leq k\leq q} f(\vx_{n+k})\right)_+\given\calD_n\right],
\end{equation}
where $y^\star_n=\min_{1\leq i \leq n} f(\vx_i)$ is the current minimum objective value, and $(z)_+ := \max(z, 0)$. 

\textbf{Information-based criteria (EIG, FI):} 
Information-based criteria select inputs that maximize information about quantities of interest, such as GP length-scales or a black-box function's global minimum. 
This information is typically measured via negative entropy, yielding the expected information gain (EIG) for Bayesian adaptive designs \citep{rainforth2024bed}, or via Fisher information \citep[FI,][]{gramacy2015local}.
Both criteria naturally extend to batch sequential settings (details in Supplement Section B).

\textbf{Space-filling criteria (MED, MaxPro):} The minimum-energy design \citep[MED,][]{joseph2015sequential} views points as charged particles in a box and minimizes the total electric potential energy.
Maximum-projection designs \citep[MaxPro,][]{joseph2015maxpro} are space-filling designs with better low-dimensional projections than Latin hypercube designs. 
Both criteria can be extended to batch sequential versions (details in Supplement Section C).

\subsubsection{Computational issues}
While extending the acquisitions above to batches is conceptually straightforward, their optimization faces computational issues.
Batch EI (\ref{eq:batch-EI}), EIG, and FI (in Supplement) involve intractable integrals over $\vy_b$, which is defined immediately after \eqref{eq:batch-ALC}. Evaluating these integrals generally requires expensive approximations.
Specifically, $q$-EI has an analytic form only for $q\leq 2$ \citep{ginsbourger2010kriging}, while EIG and FI face significant computational issues even in fully sequential settings \citep{rainforth2024bed}.
Although batch versions of ALM (\ref{eq:batch-ALM}), ALC (\ref{eq:batch-ALC}), MED, and MaxPro (in Supplement) may have closed forms, jointly optimizing them over $q$ design points remains challenging.
The resulting objective is defined on a $qd$-dimensional space and is typically nonconvex, with many local optima due to the inter-relationships among batch points.
In practice, multi-start strategies are often required, and each restart involves repeated evaluations of the batch criterion and its gradient.
These costs grow quickly with the batch size $q$ and the input dimension $d$.

\section{Methodology}\label{sec:approach}

Instead of directly optimizing a batch acquisition function $a_n(\mX_b)$, we \textit{sample} the batch from a target distribution encoding the single-point acquisition $a_n(\vx)$, which is assumed to be standardized to a common scale.
Specifically, we define the target density
\begin{equation}\label{eq:target-distribution}
    p_n(\vx) = \frac{\varphi(a_n(\vx))}{Z_n},
\end{equation}
where $\varphi$ is a positive, differentiable, and strictly increasing scalar function, and $Z_n=\int_{\bbR^d} \varphi(a_n(\vx))\diff \vx$ is the normalizing constant.
Throughout this paper, we adopt $\varphi(\cdot)=\exp(\cdot)$.
Because $\varphi$ is strictly increasing, regions of high probability align with regions of high acquisition value.
Consequently, samples drawn from $p_n$ tend to have high utilities.
However, an effective batch design also requires enough spread in the design space to reduce redundancy.
To balance these two objectives, we employ SVGD, which was introduced in Section~\ref{sec:intro}, to sample from $p_n$. 
Unlike MCMC, SVGD is a deterministic, particle-based method that updates a batch of sample points simultaneously to an equilibrium.

We give a brief review of SVGD below. 
Readers are referred to \citet{anastasiou2023stein} for a recent review on related methods.
SVGD can be viewed as a steepest descent method to reduce the KL divergence between the current density and the target density, denoted as $\varrho$ and $p$, respectively.
Consider the transport map $T(\vx) = \vx + \epsilon \bphi(\vx)$,
where $\epsilon>0$ is the step size, and $\bphi : \bbR^d \to \bbR^d$ is a vector field.
Denote $T_{\#}\varrho$ as the push forward density of $\varrho$ under $T$.
\citet{liu2016svgd} showed that, 
when $\bphi$ is restricted within the unit ball of the product reproducing kernel Hilbert space (RKHS) induced by some kernel $k$, 
the KL divergence between $T_{\#}\varrho$ and $p$ attains a maximum decreasing rate at the velocity field
\begin{equation}\label{eq:svgd-direction}
    \bphi^\star(\cdot)
    \propto
    \bbE_{\vx \sim \varrho}
    \left[
        k(\vx, \cdot)\,\nabla \log p(\vx)
        +
        \nabla_\vx k(\vx, \cdot)
    \right],
\end{equation}
and the corresponding rate can be interpreted as the \textit{kernelized Stein discrepancy} (KSD) between $\varrho$ and $p$.
Based on the above result, \citet{liu2016svgd} proposed the SVGD algorithm, which iteratively transports sample points using (\ref{eq:svgd-direction}).

In our batch design problem, we set $p$ to be the target distribution $p_n$ in (\ref{eq:target-distribution}), and the corresponding SVGD update is
\begin{equation}\label{eq:SVGD}
    \vx^{t + 1}_i
    =
    \vx^t_i
    +
    \frac{\epsilon_t}{q}
    \sum_{j=n+1}^{n+q}
    \left(
        \underbrace{
        k(\vx^t_j,\vx^t_i)\,\nabla \log p_n(\vx^t_j)
        }_{\text{driving force}}
        +
        \underbrace{
        \nabla_{\vx^t_j} k(\vx^t_j,\vx^t_i)
        }_{\text{repulsive force}}
    \right),
\end{equation}
for $i=n+1,\ldots,n+q$, where $\epsilon_t$ is the step size.
The first and second terms inside the parentheses in (\ref{eq:SVGD}) are called the \textit{driving force} and the \textit{repulsive force}, respectively.
SVGD enjoys several desired properties for batch sequential designs.
First, SVGD automatically \textit{balances} utility maximization and sample diversity.
As can be seen from the update rule (\ref{eq:SVGD}), its first term drives points toward high-probability regions, which encourages high utilities, while its second term acts as a repulsive force to push points away from each other, thereby enforcing diversity.
In the special case of $q=1$, only the driving force is present, and the point degenerates to the local optima of $p_n$ or $a_n$, reducing to the standard fully sequential design.
Second, SVGD is computationally efficient even for large $q$. 
The update rule (\ref{eq:SVGD}) has a closed form independent of the normalizing constant \(Z_n\), as $\nabla\log p_n(\vx) = \nabla\log\varphi(a_n(\vx))$.
The per-iteration cost of SVGD is $\mathcal{O}(q^2d + q c_n)$, where $c_n$ denotes the cost of evaluating $\nabla\log\varphi(a_n(\vx))$, and $\mathcal{O}(\cdot)$ is the big-$\mathcal{O}$ notation.
For \(\varphi(a)=\exp(a)\), the log gradient further reduces to \(\nabla a_n(\vx)\). 


Alternatively, other diversity-aware representative point methods can be leveraged to sample $\mX_b$ from $p_n$,
including Stein points \citep{chenWY2018stein} and support points \citep{mak2018support}.
However, Stein points typically rely on a greedy algorithm, which requires solving a series of optimization problems;
support points require estimating the unknown normalizing constant $Z_n$.
Therefore, both methods can be less efficient than SVGD.

\subsection{Quantile SVGD}
Although SVGD inherently promotes diversity, it may put points in relatively low-density (i.e., low-utility) regions.
This is undesirable when the experiments are very expensive.
One approach is to use a variant of SVGD called quantile SVGD \citep{gong2019quantile}, which was originally proposed for batch Bayesian optimization.
The key idea of quantile SVGD is to use a quantile-distorted expectation, which is sensitive to points with lower densities.
Following a similar steepest descent argument to SVGD, quantile SVGD has a closed-form update rule. 
By taking $p$ to be $p_n$ in (\ref{eq:target-distribution}) for the batch design problem and slightly adapting the original notation to the sampling-based framework, the update rule of quantile SVGD is
\begin{equation}\label{eq:qSVGD}
    \vx_i^{t+1}
    =
    \vx_i^t
    +
    \frac{\epsilon_t}{q}
    \sum_{j=n+1}^{n+q}
    \left[
        \widehat\zeta_j^t\times
        k(\vx_j^t,\vx_i^t)
        \nabla\log p_n(\vx_j^t)
        +
        \nabla_{\vx_j^t}k(\vx_j^t,\vx_i^t)
    \right],
\end{equation}
for $i=n+1,\ldots,n+q$, with
\[
    \widehat\zeta_j^t
    =
    \left(\widehat \beta_j^t\right)^{-\lambda},
    \qquad 
    \widehat \beta_j^t
    =
    \frac{1}{q}
    \sum_{\ell=n+1}^{n+q}
    \bbI\!\left\{
        \log p_n(\vx_\ell^t)
        \le
        \log p_n(\vx_j^t)
    \right\}.
\]
Here, \(\widehat \beta_j^t\) is the empirical CDF of \(\log p_n(\vx)\) at $\vx_j^t$, and $\lambda\geq 0$ is the risk-aversion parameter in the quantile-distorted expectation.
The update rule of quantile SVGD (\ref{eq:qSVGD}) differs from that of SVGD (\ref{eq:SVGD}) only in the additional weights \(\widehat\zeta_j^t\) multiplied to the driving force from point $\vx_j^t$.
When $\lambda=0$, all weights are equal to $1$, and the quantile SVGD reduces to SVGD.
When $\lambda>0$, more weights are assigned to points with lower densities.
Thus, quantile SVGD preserves the diversity-promoting repulsive force of SVGD,
while introducing risk aversion and improving the worst performance among sample points.

\subsection{Design over a constrained region}
\label{sec:constraint}

In many real-world applications, design variables are subject to physical constraints. 
For example, optimizing a Philips television tube requires minimizing weight and stress under geometric constraints on thickness and height \citep{stinstra2003constrainedMaximin}.
Similarly, the welded beam design problem \citep{deb1991optimal} and the NASA speed reducer design problem \citep{liu2017constrained} minimize the fabricating cost and the gear reducer weight, respectively, subject to various mechanical constraints like stress, deflection, and buckling load.
To restrict variables within a feasible region $\Omega \subset \bbR^d$, we consider a truncated target 
\begin{equation}\label{eq:trun-target-distribution}
    p_n(\vx)\propto \varphi\left(a_n(\vx)\right) \bbI\{\vx\in\Omega\}.
\end{equation}
This truncation is also useful when the unconstrained density \eqref{eq:target-distribution} is improper.
For example, the single-point ALM in (\ref{eq:batch-ALM}) with a stationary GP approaches the prior variance when far from the data, leading to an infinite normalizing constant unless the region is bounded. 

However, the truncated density $p_n$ is incompatible with SVGD because it is non-differentiable at the boundary and has an undefined log gradient outside $\Omega$.
Existing SVGD extensions mainly handle moment or equality constraints \citep{liu2021trustworthy,zhang2022osvgd}, while computer experiments often involve inequality constraints.
A reparameterization approach is often impractical for general inequality constraints and may encourage diversity in the transformed, rather than original, design space \citep{shi2022mirrored}.



Instead, we adapt the penalty method from constrained optimization \citep{bertsekas2014constrained}, which was recently applied to sampling by \citet{gurbuzbalaban2024penalized}. 
Assume that the design region can be represented as $\Omega=\{\vx\in\bbR^d:g_j(\vx)\le 0,\ j=1,\ldots,m\}$,
where the constraint functions $g_1,\ldots,g_m$ are differentiable. For $\mu>0$, define the penalized acquisition function $a_{n,\mu}(\vx) = a_n(\vx)-\mu\psi(\vx)$ with a quadratic penalty $\psi(\vx)=\sum_{j=1}^m (g_j(\vx))_+^2$, which remains $0$ within $\Omega$, and increases as $\vx$ deviates from $\Omega$.
The penalty parameter $\mu$ controls the overall penalty degree.
The penalized distribution 
\begin{equation}\label{eq:target-distribution-penalty}
    p_{n,\mu}(\vx)\propto\varphi(a_{n,\mu}(\vx))
\end{equation}
converges pointwise to the truncated density (\ref{eq:trun-target-distribution}) as $\mu\uparrow\infty$ (assuming that \(\varphi=\exp\)).
Because $\log p_{n,\mu}(\vx)$ is differentiable, we can apply SVGD using its gradient
$\nabla\log p_{n,\mu}(\vx) = \varphi'(a_{n,\mu}(\vx))(\nabla a_n(\vx)-\mu\nabla\psi(\vx))/\varphi(a_{n,\mu}(\vx))$,
where $\nabla\psi(\vx) = 2\sum_{j=1}^m (g_j(\vx))_+\nabla g_j(\vx)$.
Inside $\Omega$, $\nabla\psi(\vx)=0$, leaving the SVGD driving force unchanged.
Outside $\Omega$, the additional term $-\mu\nabla\psi(\vx)$ pushes points back toward $\Omega$.
The penalty increases the per-iteration cost of SVGD to $\mathcal{O}(q^2d + q c_n + q c'_m)$, with $c'_m$ being the cost of evaluating $\nabla\psi(\vx)$.

Given $\Omega$, the constraints $g_1,\ldots,g_m$ and $\varphi$ may need to be properly chosen to make $p_{n,\mu}$ a valid density.
For a common box constraint $\Omega=\prod_{i=1}^d[l_i,u_i]$, we may take 
$\{g_j(\vx)\}_{j=1}^{2d}
    =
    \{l_i-x_i\}_{i=1}^d
    \cup
    \{x_i-u_i\}_{i=1}^d$, leading to the penalty 
\begin{equation}\label{eq:penalty-box}
    \psi(\vx)
    =
    \sum_{i=1}^d
    \left\{
        (l_i-x_i)_+^2
        +
        (x_i-u_i)_+^2
    \right\},
\end{equation}
which is the squared Euclidean distance from $\vx$ to the box. 
As $\|\vx\|\rightarrow \infty$, there exists a constant $c>0$ such that
$\psi(\vx)\ge c\|\vx\|^2$. 
If $a_n$ is bounded above and $\varphi=\exp$, then $\varphi(a_{n,\mu}(\vx)) = \mathcal{O}\left(\exp(-\mu c\|\vx\|^2)\right)$, which is guaranteed to have a finite integral.

In practice, we initialize the design points inside $\Omega$ and apply SVGD to the penalized density $p_{n,\mu}$ with a sufficiently large $\mu$.
If a single large $\mu$ causes instability, we can gradually increase $\mu$, and repeatedly apply SVGD to each $\mu$ using warm starts from the previous SVGD results.

\subsection{Local Penalty Regularization}
\label{sec:local-penalty}
Although the repulsive force in SVGD promotes diversity within the newly selected batch, it does not repel them from previously selected points.
Such a separation is induced through the driving force, that is, through the target density or the acquisition function $a_n$, which is typically small near the existing inputs.
However, when the target density is nearly flat, the driving force becomes weak over most of the design region, so SVGD may diversify the new batch points among themselves without ensuring separation from the existing ones.
This can lead to redundant selections, motivating the additional regularization below.

Let $\vx_1,\ldots,\vx_n$ denote the existing design points.
We augment the design region $\Omega$ with $n$ minimum-distance constraints
$\{\norm{\vx - \vx_i}\geq r_i\}$, for $i=1,\ldots,n$. 
Equivalently, the $i$-th constraint excludes the ball of radius $r_i$ centered at $\vx_i$.
We formulate the $i$-th constraint as $h_i(\vx) = 1 - \norm{\vx - \vx_i}^2/r_i^2 \leq 0$, which is differentiable.
Following the same idea in Section~\ref{sec:constraint}, we define the locally penalized acquisition function
\begin{equation*}
  a_{n,\mu,\kappa}(\vx) = a_n(\vx) - \mu\,\psi(\vx) - \kappa\,\xi(\vx), \quad \psi(\vx)=\sum_{j=1}^m (g_j(\vx))_+^2,
      \quad \xi(\vx) = \sum_{i=1}^n (h_i(\vx))_+^2,
\end{equation*}
and the corresponding target density
\begin{equation}\label{eq:target-distribution-local-penalty}
    p_{n,\mu,\kappa}(\vx) \propto \exp\!\left(a_{n,\mu,\kappa}(\vx)\right).
\end{equation}
Here, $\kappa\ge0$, similar to $\mu$, is a penalty parameter controlling the strength of the local penalty constraints.
Outside all local balls, $\xi(\vx)=0$, so the target density is unchanged.
Inside the ball around $\vx_i$, the gradient 
$-\kappa\nabla\xi(\vx) = 4\kappa \sum_{i=1}^n (\vx-\vx_i)(h_i(\vx))_+/r_i^2$
pushes particles away from $\vx_i$.
Thus, the global penalty keeps particles inside the feasible region, while the local penalty keeps them away from the previous points.
The additional evaluation of $\nabla\xi(\vx)$ costs $\mathcal{O}(nd)$, which is typically no more expensive than evaluating $\nabla a_n(\vx)$.
As a result, the overall scalability of the proposed method is not affected.
We refer to this modification as the \textit{local penalty} method, due to its similarity with an existing approach in \citet{gonzalez2016batch} for batch Bayesian optimization.
However, the latter is designed to promote within-batch diversity, whereas our local penalty serves an entirely different purpose.

The choice of radii $\{r_i\}$ is more subtle, and should depend on the design problem.
Larger radii enforce stronger separation from existing points and lead to a more space-filling design, but overly large radii may exclude high-utility regions, making the batch driven mainly by separation rather than by the acquisition function. 
On the other hand, smaller radii impose weaker exclusion around existing points, so new points may still be placed in a previously explored region when further sampling there is valuable.
Specific choices of $\{r_i\}$ will be given in Section~\ref{sec:numerical} for two fully sequential methods.
The overall sampling-based batch sequential design is summarized in Algorithm~\ref{alg:svgd-batch}.

\begin{algorithm}[!htb]
\caption{Sampling-based batch sequential design via SVGD}
\label{alg:svgd-batch}
\begin{algorithmic}[1]
\Require Initial data $\calD_{n_0}$, batch size $q$, batch number $B$, fully sequential acquisition function $a_n$, constraint functions $\{g_j\}_{j=1}^m$, penalty parameter $\mu$, SVGD iteration number $T$, base kernel $k$, SVGD step sizes, SVGD initialization scheme, 
\Optional Risk-aversion parameter $\lambda$, local penalty parameter $\kappa$, local radii rule.
\For{$n=n_0,n_0+q,\ldots,n_0+(B-1)q$}
    \State Fit the surrogate model based on $\calD_n$.
    \State Construct the acquisition function $a_n$, and standardize it to a common scale.
    \State Initialize points $\{\vx_{n+1}^{(0)},\ldots,\vx_{n+q}^{(0)}\}$ in $\Omega$.
    \For{$t=0,\ldots,T-1$}
        \State \parbox[t]{0.82\linewidth}{Update the points $\{\vx_{n+1}^{(t)},\ldots,\vx_{n+q}^{(t)}\}$ by SVGD \eqref{eq:SVGD} or quantile SVGD \eqref{eq:qSVGD}, with $p_n$ replaced by $p_{n,\mu}$ (\ref{eq:target-distribution-penalty}, without local penalty) or $p_{n,\mu,\kappa}$ (\ref{eq:target-distribution-local-penalty}, with local penalty).}
    \EndFor
    \State Set $\mX_b^*=\{\vx_{n+1}^{(T)},\ldots,\vx_{n+q}^{(T)}\}$, optionally projected onto $\Omega$.
    \State Evaluate the responses $\vy_b^*$.
    \State Update $\calD_{n+q} = \calD_n\cup\{(\mX_b^*,\vy_b^*)\}$.
\EndFor
\State \textbf{return} $\calD_{n_0+Bq}$.
\end{algorithmic}
\end{algorithm}

\section{Case studies and numerical results}\label{sec:numerical}

We apply the proposed sampling-based framework to obtain batch versions of ALM using two different surrogate models. 
While ALM is used for illustration, our framework applies to any fully sequential method with a differentiable acquisition function.
In \cref{sec:ALM-stationary}, we model $f$ as a stationary GP to demonstrate a near-uniform target distribution and the validity of the local penalty method.
In \cref{sec:ALM-nonstationary}, we use a state-of-the-art nonstationary GP, showing that our approach achieves comparable performance to fully sequential design and the batch design by direct extension at a significantly lower cost.

For the acquisition function, let $s_n^2(\vx) = \bbV\!\left(f(\vx)\given\calD_n\right)$ denote the plug-in posterior variance of $f(\vx)$, with all unknown parameters estimated by MLE.
To handle different models and test functions, we rescale $s_n^2$ to a common scale:
$a_n(\vx) = s_n^2(\vx)/c_n$, where $c_n$ is the running maximum of the $95\%$ quantile of $s_n^2$ on $1000d$ Sobol' points \citep{santner2018design}.
This running maximum is found to be more stable than the $95\%$ quantile of the current $s_n^2$.
The same Sobol' set is also used to evaluate the prediction performance.


Throughout this section, we consider a box constraint $\Omega = [0,1]^d$.
For the proposed Algorithm~\ref{alg:svgd-batch}, we use quantile SVGD with the risk-aversion parameter $\lambda = 1.5$, as it improves the worst-case performance relative to standard SVGD.
The penalty function is \eqref{eq:penalty-box}, with penalty parameter $\mu = 10^4$. If the local penalty method is used, we set $\kappa = 1$ to produce soft minimum-distance constraints.
Following \citet{liu2016svgd}, we use the RBF kernel $k(\vx,\vx') = \exp\!(-\norm{\vx-\vx'}^2/h^2)$, with $h^2 = \texttt{median}^2/\log q$ and AdaGrad step sizes.
Here, \texttt{median} is the median of pairwise distances between the current points. 
SVGD is initialized at the first $q$ Sobol' points at least $0.5\,n^{-1/d}$ away from the existing inputs.
After $10^3$ iterations, points are projected onto $\Omega$.
All methods are implemented in the R programming language on a laptop with a 2.2-GHz Intel Core i9-14900HX processor and $32$-GB RAM.

\subsection{ALM with a stationary GP}
\label{sec:ALM-stationary}

We assume an ordinary kriging (OK) model for the noiseless response:
\begin{equation}\label{eq:OK-model}
    y = f(\vx) = \mu + \nu\cdot Z(\vx), \qquad Z(\cdot) \sim \GP\!\left(0, R_{\btheta}\right),
\end{equation}
where $Z$ is a GP with mean $0$ and an anisotropic Gaussian correlation function $R_{\btheta}(\vx, \vx') = \exp\!\left(-\sum_{l=1}^{d}(x_l - x_l')^2/(2\theta_l^2)\right)$; $\btheta = \{\theta_{1}, \ldots, \theta_{d}\}$ are the per-coordinate length-scale parameters; $\nu>0$ is the global scale parameter.
Under the OK model, the posterior variance $s_n^2$ has a closed-form
\begin{equation}\label{eq:ALM-OK}
    s_n^2(\vx) = \nu^2\left(1 - \vr(\vx)\T\mR^{-1}\vr(\vx)\right),
\end{equation}
where $\vr(\vx) = \bigl(R_{\btheta}(\vx, \vx_1), \ldots, R_{\btheta}(\vx, \vx_n)\bigr)\T$ and $\mR = \bigl(R_{\btheta}(\vx_i, \vx_j)\bigr)_{i,j=1}^{n}$ are the correlation vector and matrix, respectively.
All unknown parameters are estimated by MLE and plugged in.
Starting from an initial MaxPro design with $n_0=10$ points,
we sequentially add three batches, each containing $q=10$ points.
The true $f$ is the Branin function \citep{simulationlib} with the input region rescaled to $\Omega = [0,1]^2$.

\begin{figure}[!htb]
  \centering
  \begin{subfigure}{\textwidth}
      \includegraphics[height=0.28\textwidth]{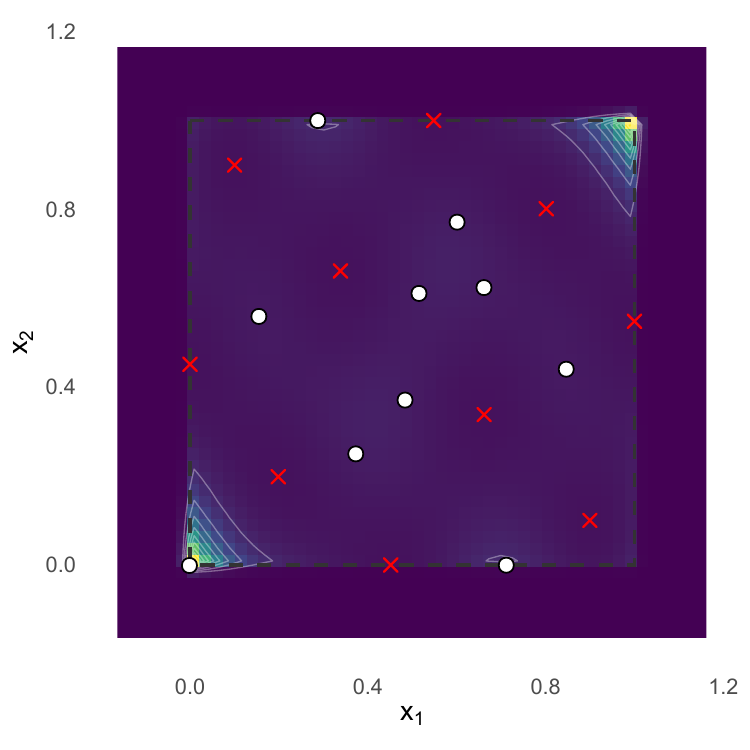}\hfill
      \includegraphics[height=0.28\textwidth]{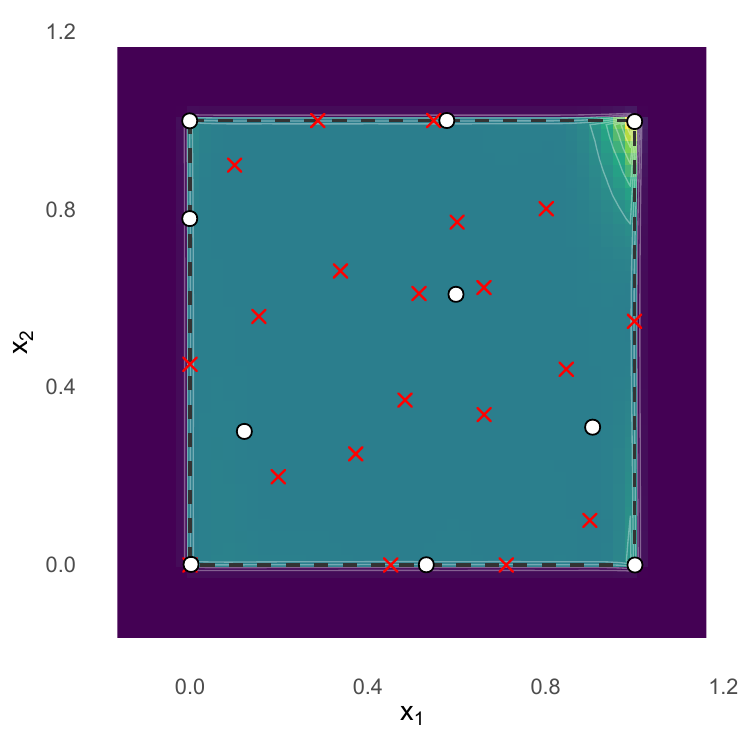}\hfill
      \includegraphics[height=0.28\textwidth]{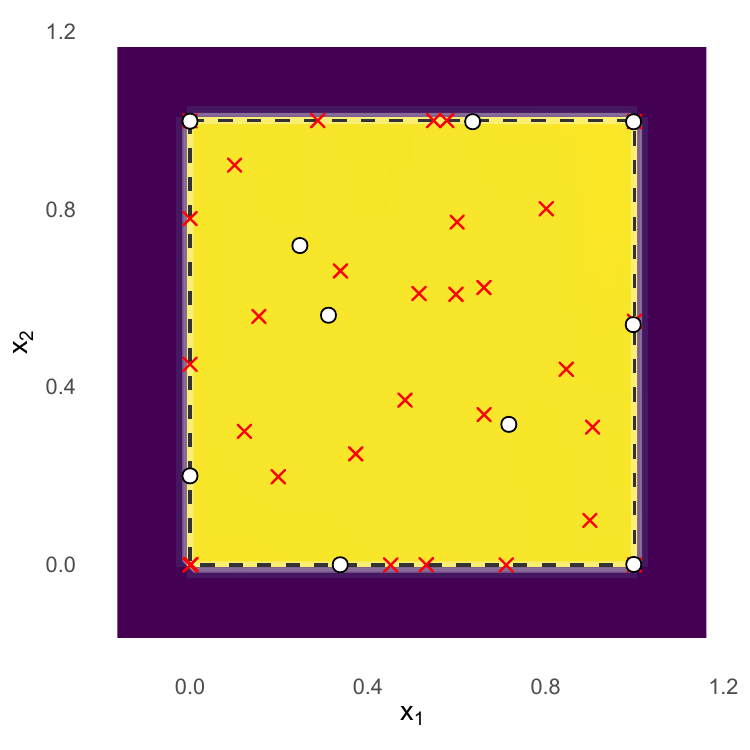}\hfill
      \includegraphics[height=0.28\textwidth]{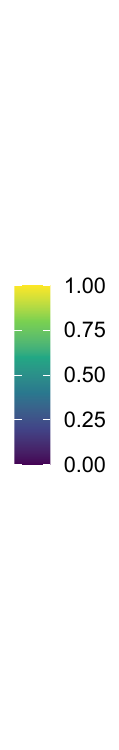}
      \caption{Batches $1-3$ without the local penalty}
      \label{fig:branin-rounds-without-local-penalty}
  \end{subfigure}\\[1.5ex]
  \begin{subfigure}{\textwidth}
      \includegraphics[height=0.28\textwidth]{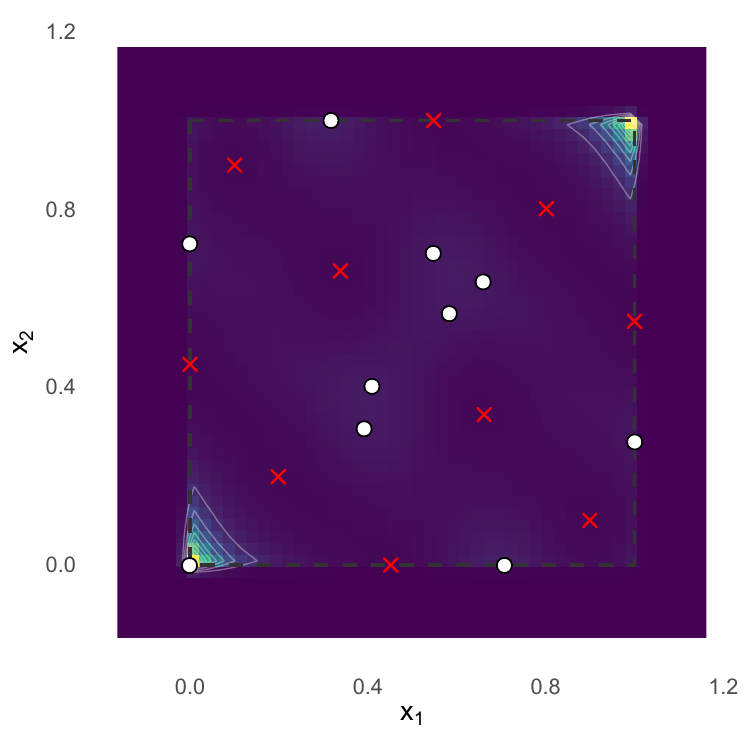}\hfill
      \includegraphics[height=0.28\textwidth]{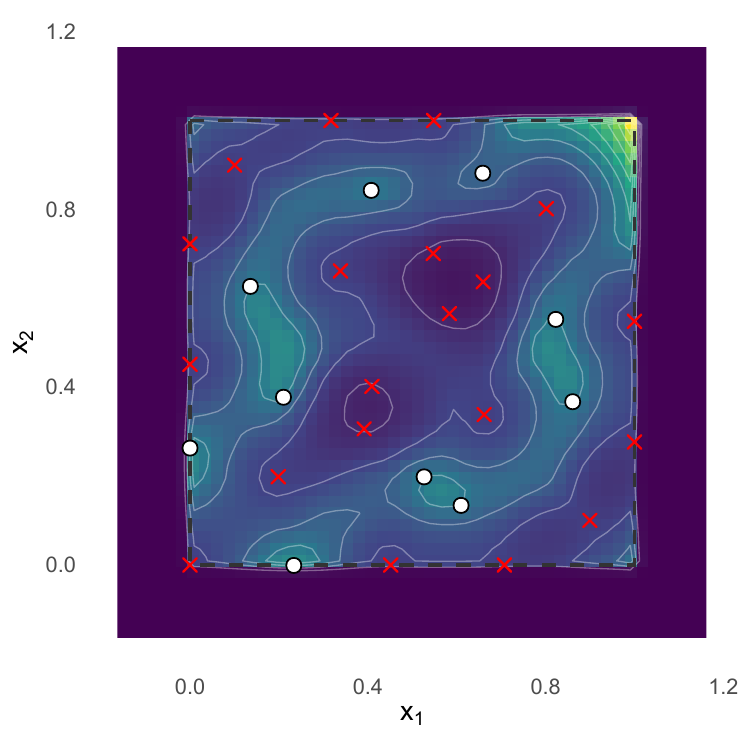}\hfill
      \includegraphics[height=0.28\textwidth]{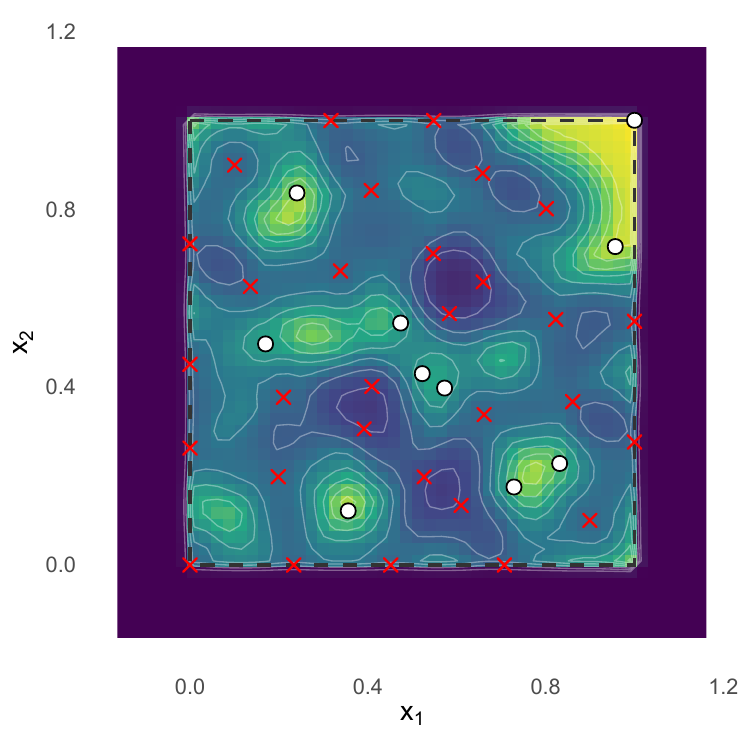}\hfill
      \includegraphics[height=0.28\textwidth]{figures/colorbar.pdf}
      \caption{Batches $1-3$ with the local penalty}
      \label{fig:branin-rounds-with-local-penalty}
  \end{subfigure}
  \caption{Batches $1-3$ (from left to right) and the target density normalized to $[0,1]$ (the heatmaps) in each round. The top (resp., bottom) row uses quantile SVGD without (resp., with) the local penalty. Red crosses are the training points, and white circles are the newly selected batch of points.}
  \label{fig:branin-rounds}
\end{figure}

\begin{figure}[htbp]
  \centering
  \begin{subfigure}[t]{0.42\textwidth}
      \centering
      \includegraphics[width=0.8\textwidth]{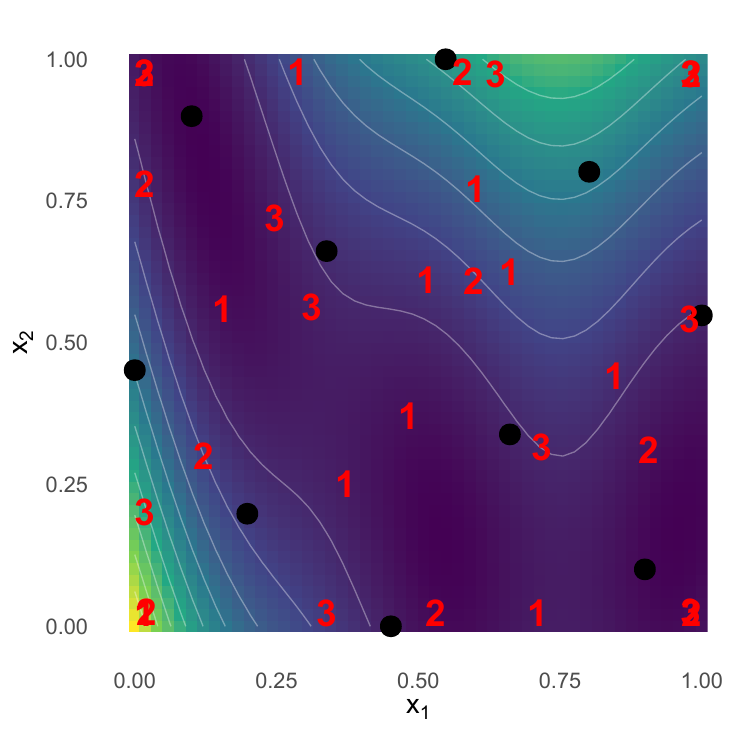}
      \caption{Batches $1-3$ without the local penalty}
      \label{fig:local-penalty-without-penalty}
  \end{subfigure}\hspace{0.2in}
  \begin{subfigure}[t]{0.42\textwidth}
      \centering
      \includegraphics[width=0.8\textwidth]{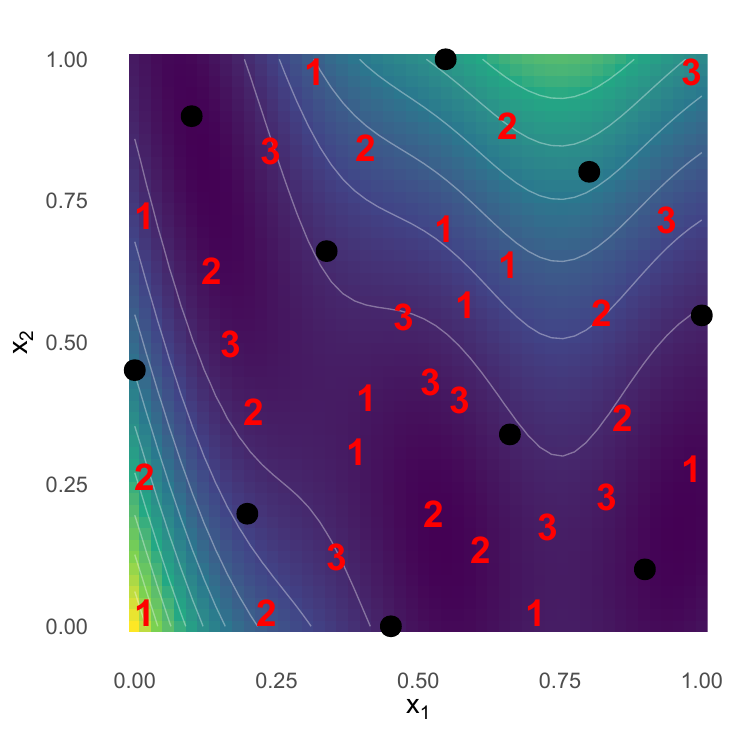}
      \caption{Batches $1-3$ with the local penalty}
      \label{fig:local-penalty-with-penalty}
  \end{subfigure}\hspace{0.1in}
  \includegraphics[height=0.4\textwidth]{figures/colorbar.pdf}
  \caption{Batches $1-3$ (red numbers) and the Branin function (heatmap, rescaled to $[0,1]$). The left (resp., right) panel uses quantile SVGD without (resp., with) the local penalty. The black dots are $n_0 = 10$ initial MaxPro designs.}
  \label{fig:local-penalty}
\end{figure}

As shown in Figures \ref{fig:branin-rounds-without-local-penalty} and \ref{fig:local-penalty-without-penalty}, without the local penalty, the SVGD points begin to overlap with existing data in batches $2$ and $3$.
The target densities are almost uniform, as can be seen from Figure \ref{fig:branin-rounds-without-local-penalty}.
One possible reason for flat densities is that the overall magnitude of posterior variance $s_n^2$ decreases as data are added.
Another possible reason is due to the stationarity of the OK model.
By construction, the training data only affects the ALM criterion (\ref{eq:ALM-OK}) via the length-scale parameters $\btheta$ (after ignoring the constant factor $\nu^2$).
The MLE of $\btheta$ can be numerically unstable on a small dataset and return extreme estimates, which exacerbate the uniform density issue.

We then test the local penalty method described in Section~\ref{sec:local-penalty}.
Regarding the local radii, because ALM with the OK model tends to produce space-filling designs \citep[see, e.g., Chapter 7 of][]{joseph2026experimental}, we adopt a uniform radius $r_i\equiv r$ for all points.
By the intuition that $n$ balls of radius $r$ scattered uniformly have a total volume of order $O(n r^d)$, we take $r = O(n^{-1/d})$.
In Section~\ref{sec:ALM-nonstationary}, we adopt data-adaptive $r_i$ to align with the nonstationary model. 
As seen in Figures \ref{fig:branin-rounds-with-local-penalty} and \ref{fig:local-penalty-with-penalty}, with the local penalty regularization, the target densities are no longer uniform, and the redundancy has been substantially reduced.

\subsection{ALM with a nonstationary GP}
\label{sec:ALM-nonstationary}
\subsubsection{The HRK model}

As mentioned in the previous subsection, variance-based criteria with a stationary GP tend to place points uniformly in the space.
In this section, we apply the sampling-based framework to convert ALM with heteroskedastic rational kriging \citep[HRK,][]{wang2026hrk} into a batch version.
HRK is based on Rational kriging \citep[RK,][]{joseph2025rational}, which models the noiseless response as
\begin{equation}\label{eq:hrk-model}
    y = f(\vx) = \mu + \tau(\vx)\cdot Z(\vx), \quad Z(\cdot) \sim \GP\!\left(0, R_{\btheta}\right), \quad
    \tau(\vx) = \frac{\nu}{c_0 + \vr(\vx)\T\vc}.
\end{equation}
As in the OK model (\ref{eq:OK-model}), $Z$ is a GP with mean $0$ and an anisotropic Gaussian correlation $R_{\btheta}$.
The key difference between OK and RK is the spatially varying scale parameter $\tau(\vx)$, which models the heterogeneity of $f$.
The specific form of $\tau(\vx)$ is motivated by the equivalence of the posterior mean of $f$ under the RK model and the best unbiased rational predictor
$\hat y(\vx) = \mu + (\vr(\vx)\T\mathbf{b})/(c_0 + \vr(\vx)\T\vc)$.
Here, the unknown parameters $\tilde\vc= (c_0, \vc\T)\T\geq 0$ are non-negative and normalized such that $\norm{\tilde\vc} = 1$.
HRK shares the same framework as RK while being different in the estimation of $\tilde\vc$.
RK assigns a non-informative prior to $\mu$ and estimates $\tilde\vc$ by minimizing the posterior variance of $\mu$, which tends to produce an almost flat $\tau(\vx)$ in some scenarios.
In comparison, HRK estimates $\tilde\vc$ by MLE, which encourages $\tau(\vx)$ to be more spatially varying.
The plug-in ALM criterion under an HRK model is given by
\begin{equation}\label{eq:hrk-variance}
    s_n^2(\vx)
    = \tau^2(\vx) \left(1 - \vr(\vx)\T\mR^{-1}\vr(\vx)\right)
    = \frac{\nu^2}{\left(c_0 + \vr(\vx)\T\vc\right)^2} \left(1 - \vr(\vx)\T\mR^{-1}\vr(\vx)\right).
\end{equation}

Following the practice introduced at the beginning of Section~\ref{sec:numerical}, we rescale the raw criterion $s_n^2(\vx)$ in (\ref{eq:hrk-variance}) by the running maximum of $95\%$ quantiles $c_n$, and obtain $a_n = s_n^2/c_n$.
The gradient has a closed form, which is given in Supplement Section A.
Compared to the stationary case, $a_n$ now incorporates both the $\left(1 - \vr(\vx)\T\mR^{-1}\vr(\vx)\right)$ term, which encourages space-filling, and the $\tau^2(\vx)$ term, which favors regions with larger fluctuations.
Such an adaptive nature motivates a non-uniform scheme for the local penalty radii $\{r_i\}$:
\begin{equation}\label{eq:hrk-radii}
    r_i \propto n^{-1/d}\,\left(\frac{\tau(\vx_i)}{\min_{1\leq j \leq n} \tau(\vx_j)}\right)^{-1}, \qquad i = 1, \ldots, n.
\end{equation}
Therefore, points in highly variable regions have smaller radii (allowing denser sampling), while points in flatter regions have larger radii (encouraging space-filling).
In particular, when $\tau(\vx)\equiv 1$, the radius (\ref{eq:hrk-radii}) reduces to the uniform $r_i \propto n^{-1/d}$ used in Section~\ref{sec:ALM-stationary}.

\subsubsection{Results}
We use the two-dimensional Gramacy--Lee test function \citep{gramacy2009adaptive} to demonstrate active learning with HRK.
The test function is defined as $f(\vx) = \tilde x_1 \exp(-\tilde x_1^2 - \tilde x_2^2)$, with $\tilde x_i = 6x_i - 2$ and $x_i\in[0,1]$ for $i=1,2$.
The Gramacy--Lee function is highly nonlinear in the lower-left corner, while being almost flat in other regions.
We generate the initial design of $n_0 = 20$ points using MaxPro and add three batches sequentially, with each batch containing $q = 10$ points.
We compare three methods. 
\begin{enumerate}[nosep]
    \item \textbf{Fully sequential ALM.} Sequentially adds $30$ points one by one by maximizing the posterior variance \eqref{eq:hrk-variance} via L-BFGS (over $d$ variables, with $10$ restarts at maximin Latin hypercube designs) and updates the model each time. The L-BFGS was implemented via the R package \texttt{nloptr}, which automatically handles box constraints.
    \item \textbf{Batch ALM by direct extension.} Following \cref{sec:extension}, the batch is selected by maximizing the log-determinant of the batch posterior covariance matrix. 
    Under HRK, with the plug-in parameters, $\Cov\!\left(\vf_b\given\calD_n\right) = \mD_b \mR_{b\mid n} \mD_b$, where $\mD_b =\operatorname{diag}\!\left\{\tau(\vx_{n+1}), \ldots, \tau(\vx_{n+q})\right\}$;
    $\mR_{b\mid n} = \mR_b - \mR_{nb}\T\mR^{-1}\mR_{nb}$;
    $\mR_b$ is the $q \times q$ correlation matrix of $\{\vx_{n+1},\ldots,\vx_{n+q}\}$;
    $\mR_{nb}$ is the $n \times q$ cross-correlation matrix with the $j$-th column as $\vr(\vx_{n+j})$, $j=1,\ldots,q$. Hence,
    \begin{equation}\label{eq:hrk-batch-logdet}
        \log\det\Cov\!\left(\vf_b\given\calD_n\right)= 2\sum_{j=1}^{q}\log\tau(\vx_{n+j}) + \log\det\mR_{b\mid n}.
    \end{equation}
    Interestingly, the batch acquisition function (\ref{eq:hrk-batch-logdet}) explicitly balances the utility and diversity.
    The first term stems from the heterogeneity of HRK, encouraging points in regions with larger fluctuations.
    The second term promotes diversity.
    The objective function in \eqref{eq:hrk-batch-logdet} is jointly maximized by using L-BFGS (over $qd$ variables).
    The closed-form gradients are given in Supplement Section A. We use $20$ restarts at maximin Latin hypercube designs. 
    \item \textbf{Batch ALM via SVGD}. The batch is selected by our proposed Algorithm~\ref{alg:svgd-batch}, using quantile SVGD, the same penalty function (\ref{eq:penalty-box}) as in Section~\ref{sec:ALM-stationary}, and local penalty with adaptive radii (\ref{eq:hrk-radii}). 
\end{enumerate}

\begin{figure}[htbp]
    \centering
    \begin{subfigure}[t]{0.3\textwidth}
        \centering
        \includegraphics[height=\textwidth]{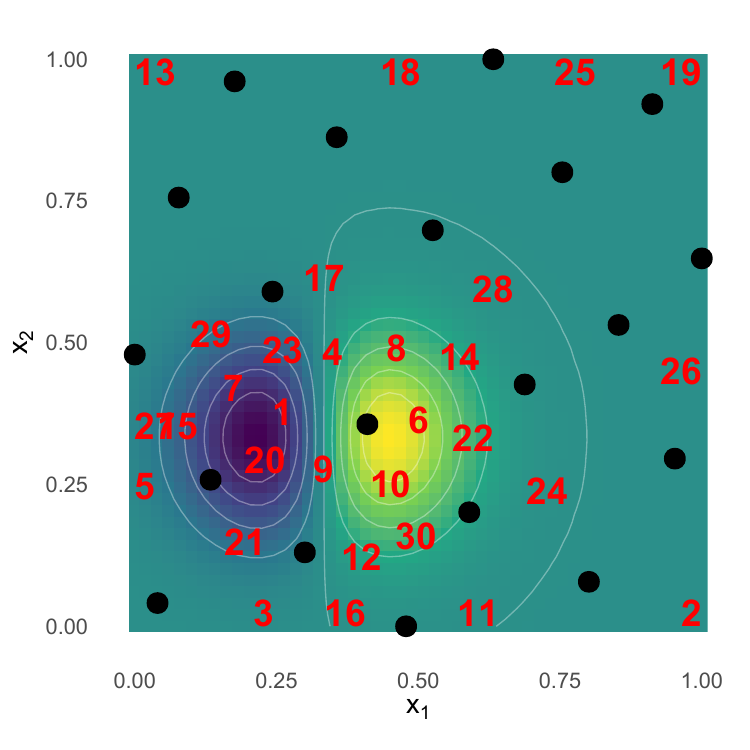}
        \caption{Fully sequential}
    \end{subfigure}
    \begin{subfigure}[t]{0.33\textwidth}
        \centering
        \includegraphics[height=0.9\textwidth]{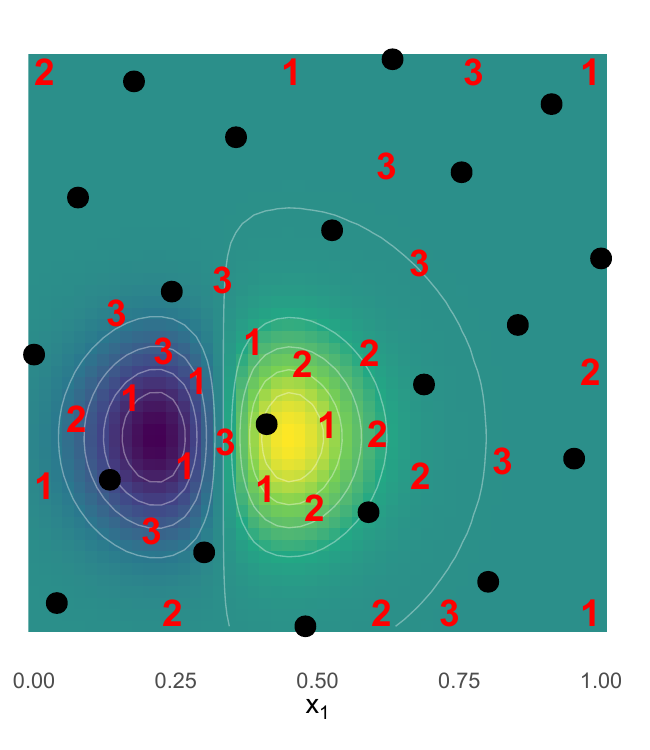}
        \caption{Batch by direct extension}
    \end{subfigure}\hspace{-.2in}
    \begin{subfigure}[t]{0.33\textwidth}
        \centering
        \includegraphics[height=0.9\textwidth]{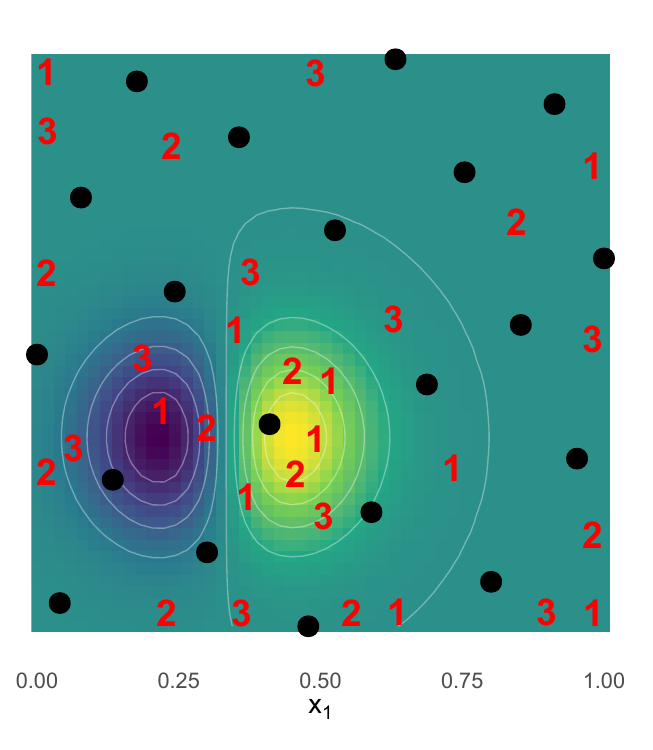}
        \caption{Batch by SVGD}
    \end{subfigure}\hspace{-.1in}
    \includegraphics[height=0.31\textwidth]{figures/colorbar.pdf}
    \caption{Final results of the three methods on the Gramacy--Lee test function (heatmap).
    Black dots are $20$ initial designs.
    Red numbers indicate the round and position for the points added ($1$--$30$ for the fully sequential design, $1$--$3$ for the batch designs).}
    \label{fig:hrk-comparison}
\end{figure}

\cref{fig:hrk-comparison} shows that all three methods tend to put more points in the lower-left corner, where $f$ is more nonlinear, and on the boundary, which is due to the nature of ALM.
It can be observed that the fully sequential method puts fewer points in the flat region than the two batch methods.
One possible reason is that the fully sequential method updates its model after every single point, allowing it to quickly identify flat areas. 
Nevertheless, the fully sequential method serves only as a benchmark here.
As discussed in Section~\ref{sec:intro}, fully sequential and batch methods are designed for different settings. 
When multiple experiments must be evaluated in parallel, the point-by-point selection is not an option.

We next test the prediction performance of the three methods on various test functions with dimensions varying from $d = 2$ to $d=8$, including the Gramacy--Lee function ($d=2$) above, the curved function of Dette and Pepelyshev ($d = 3$), the cantilever beam ($d = 4$), OTL circuit ($d = 6$), piston ($d = 7$), and borehole ($d = 8$) functions.
The analytic forms of test functions can be found in \citet{simulationlib}.
All methods start with the same $20$ initial design points by MaxPro.
The fully sequential method adds one point at a time, for a total of $80$ points.
The batch methods add $q=10$ points in each iteration, for a total of $8$ batches.
We additionally test SVGD with no local penalty as a comparison.
The experiment is repeated for $10$ times due to the randomness in generating the initial MaxPro design.
A test set of $1000d$ Sobol' points is used to evaluate the prediction performance. 

\begin{figure}[!htb]
    \centering
    \begin{subfigure}{0.28\textwidth}
        \includegraphics[width=\textwidth]{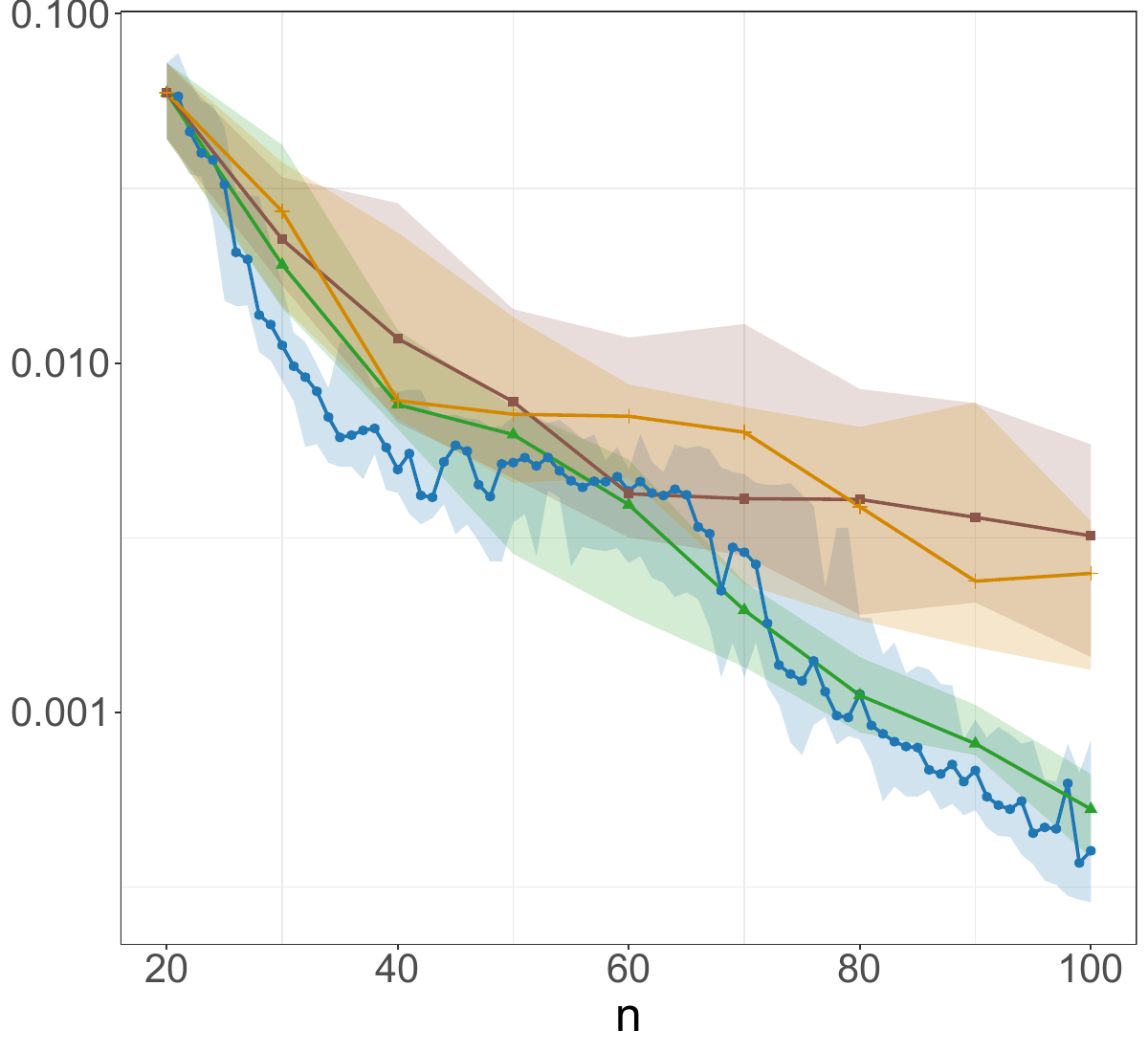}
        \caption{Gramacy--Lee}
    \end{subfigure}\hfill
    \begin{subfigure}{0.28\textwidth}
        \includegraphics[width=\textwidth]{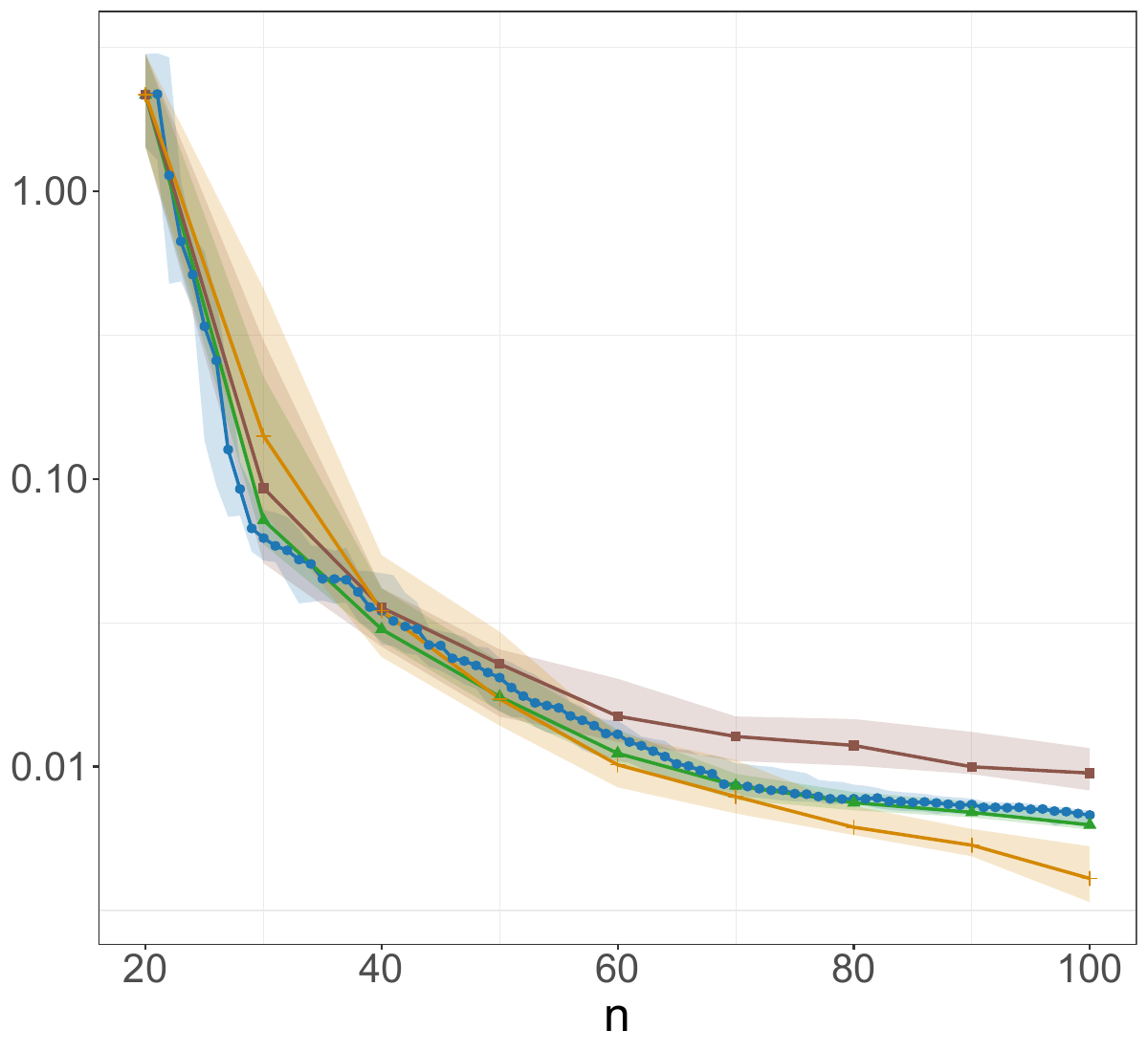}
        \caption{Dette--Pepelyshev}
    \end{subfigure}\hfill
    \begin{subfigure}{0.28\textwidth}
        \includegraphics[width=\textwidth]{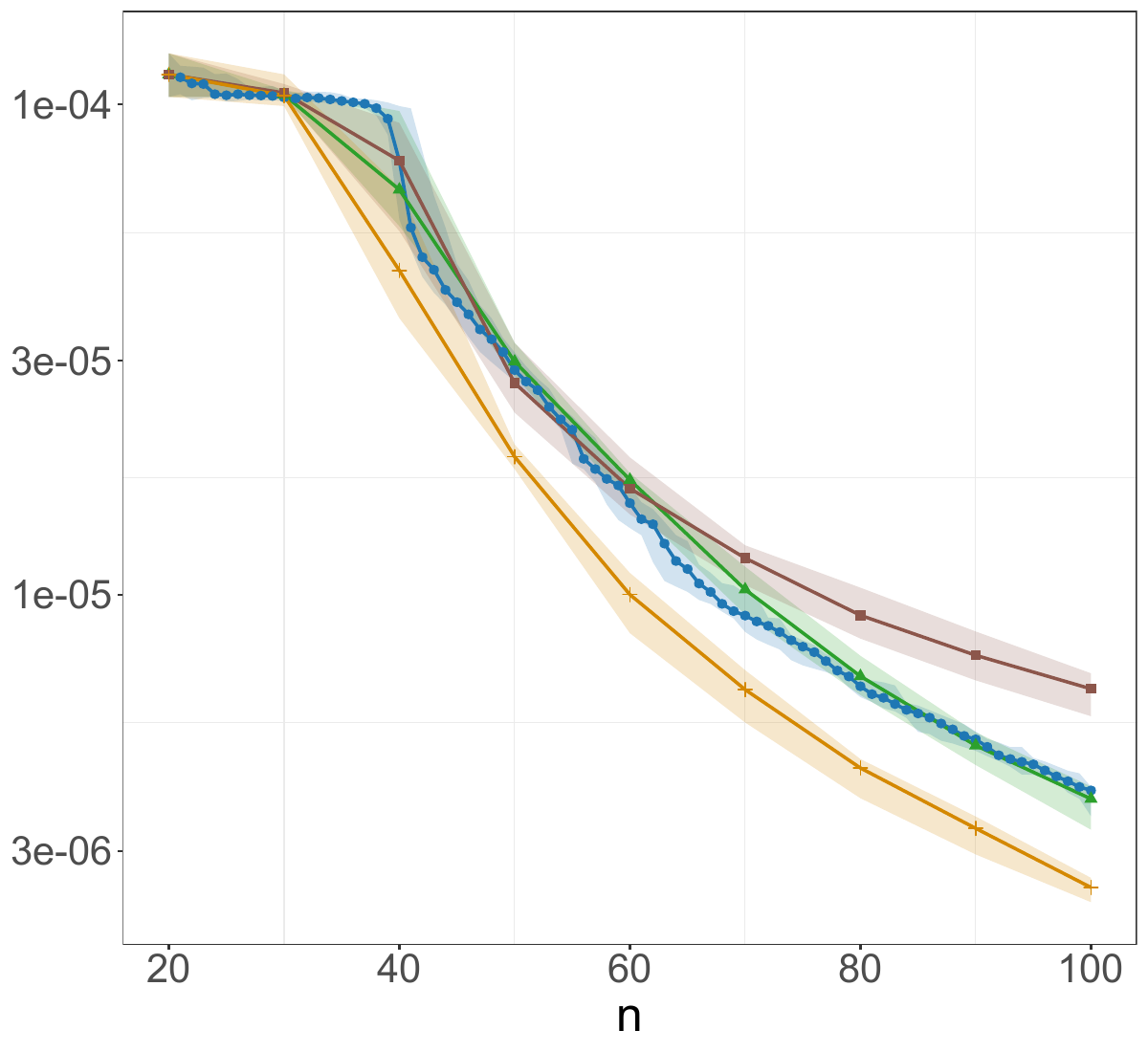}
        \caption{Cantilever beam}
    \end{subfigure}\\[1.5ex]
    \begin{subfigure}{0.28\textwidth}
        \includegraphics[width=\textwidth]{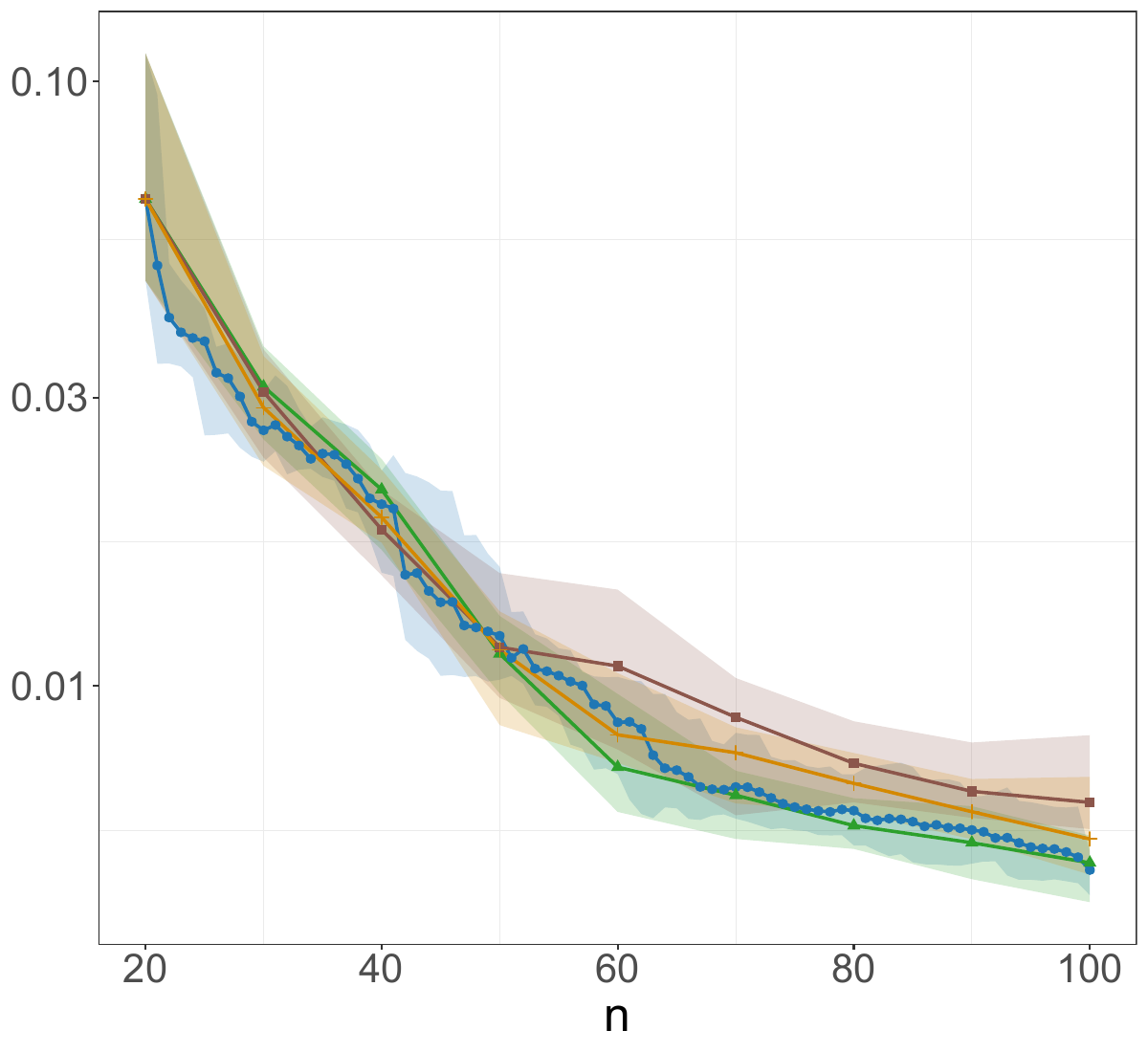}
        \caption{OTL circuit}
    \end{subfigure}\hfill
    \begin{subfigure}{0.28\textwidth}
        \includegraphics[width=\textwidth]{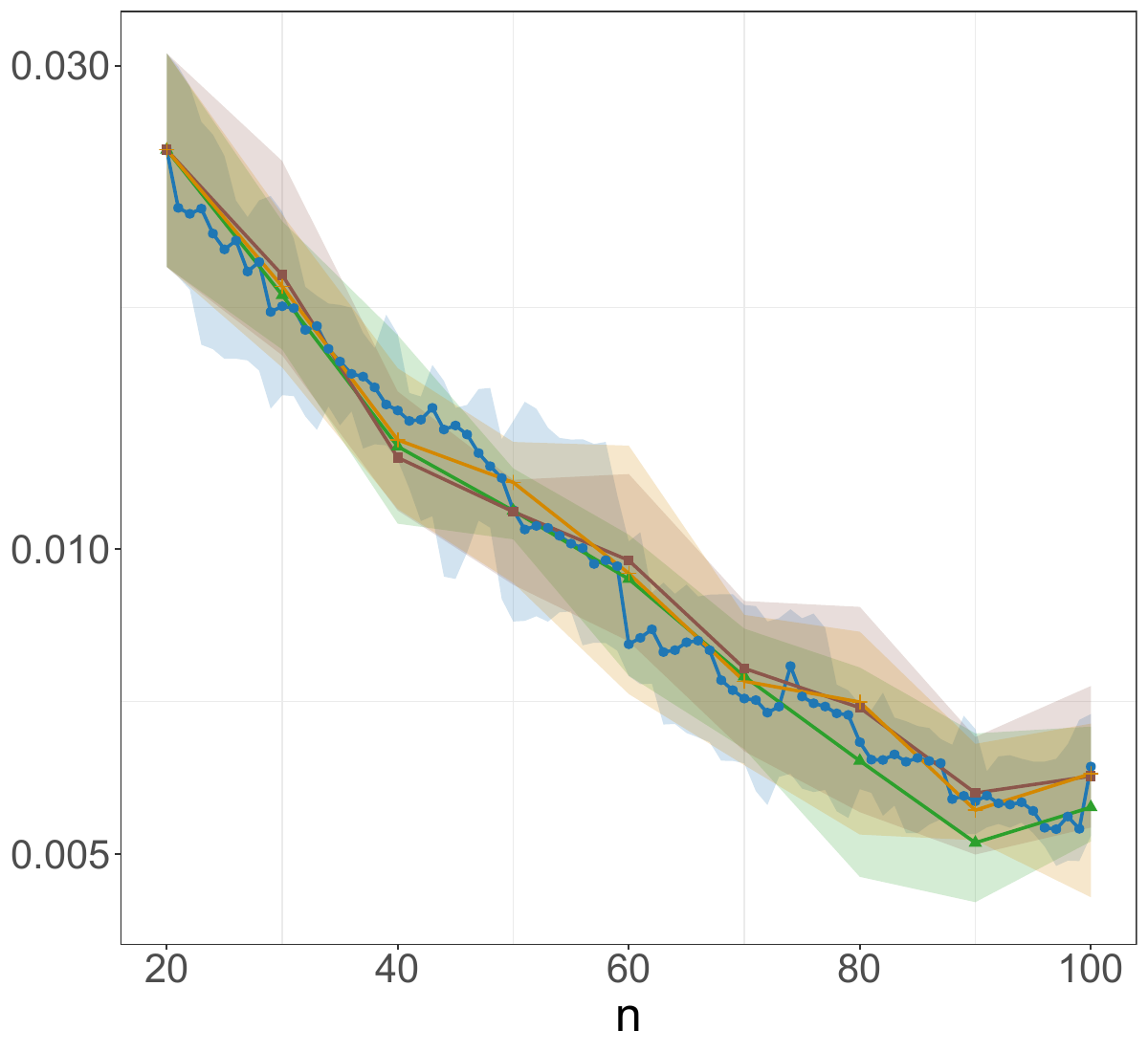}
        \caption{Piston}
    \end{subfigure}\hfill
    \begin{subfigure}{0.28\textwidth}
        \includegraphics[width=\textwidth]{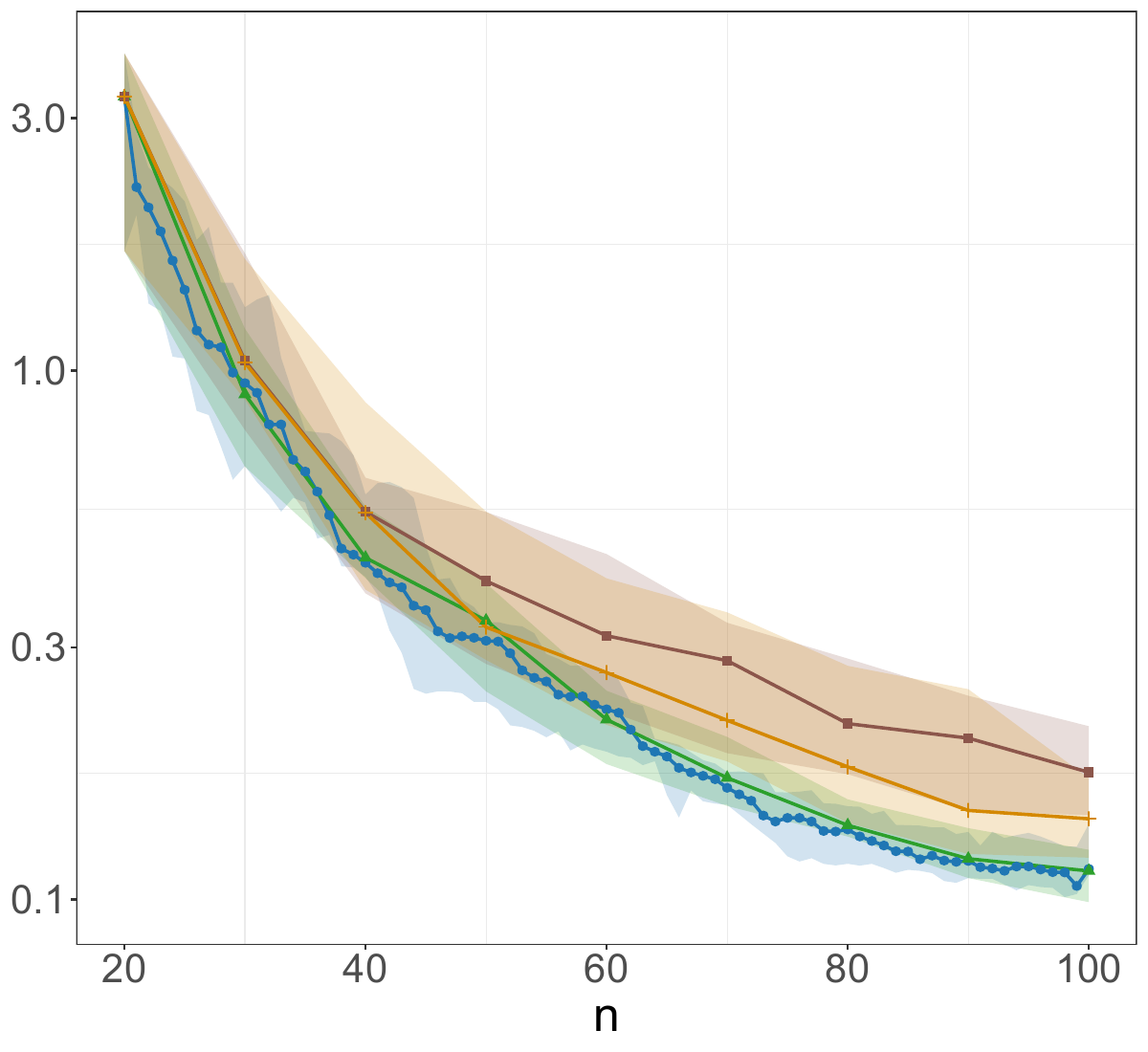}
        \caption{Borehole}
    \end{subfigure}\\[2ex]
    \includegraphics[width=0.7\textwidth]{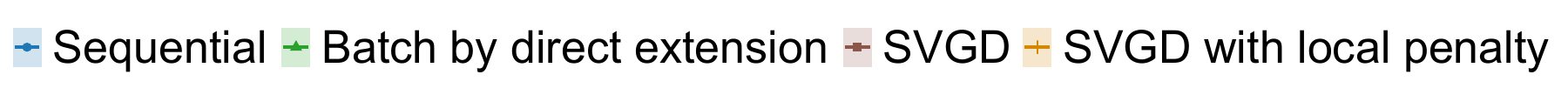}
    \caption{
    Prediction performance (in RMSE) for fully sequential and batch sequential methods. Solid lines represent the median over $10$ repetitions, and the shaded bands mark the $10$th and $90$th quantiles.}
    \label{fig:hrk-rmse-benchmark}
\end{figure}

\cref{fig:hrk-rmse-benchmark} reports the prediction performance (RMSE) of four methods in $10$ repetitions. 
Use of the local penalty tends to improve the performance of SVGD as the total number of design points increases, which indicates that the near-uniform density issue persists even with nonstationary GPs.
Moreover, SVGD with the local penalty is comparable to the fully sequential and batch-by-definition methods.
It even outperforms both baselines on the Dette--Pepelyshev and Cantilever beam functions.
One possible reason is that these two surfaces are less heterogeneous, and SVGD with the local penalty promotes diversity better, which makes it more efficient.
Our method only struggles slightly on the Gramacy--Lee function, likely because the latter is highly nonstationary, so that the acquisition landscape becomes multi-modal with many local maxima, which is difficult to sample from.

Finally, we compare the processing time of batch by direct extension and batch by SVGD with the local penalty. 
Batch by direct extension requires optimizing \eqref{eq:hrk-batch-logdet} over $qd$ variables with multiple restarts, whereas each SVGD iteration has a closed-form update and only requires $q$ gradient evaluations of the single-point acquisition function.
Details are deferred to the Supplement.
Overall, as the batch size increases, batch by SVGD takes substantially less time than batch by direct extension.


\section{Summary and Further Discussion}\label{sec:conclusion}
In this article, we propose a sampling-based framework to systematically convert any fully sequential experimental design with a differentiable acquisition function into a batch sequential design.
Instead of performing computationally expensive joint optimization over multiple design points, our approach transforms batch selection into a sampling problem.
We utilize Stein variational gradient descent (SVGD) to draw samples from a target density that encodes the single-point utility, inherently balancing the selection of high-utility points with the overall diversity for effective batches.

To address the new challenges of applying SVGD to experimental designs, we adapt the quadratic penalty methods. 
First, we use a global penalty to restrict SVGD updates within constrained design spaces. 
Second, we develop an adaptive local penalty method to prevent overlapping when the target distribution becomes nearly uniform. 
Our numerical studies, based on both stationary and nonstationary surrogate models, demonstrate that this sampling-based framework achieves predictive performance comparable to standard batch-by-definition methods, while being computationally more efficient and scalable with respect to batch size.

A direction for future research is to extend this framework to accommodate mixed-type (e.g., continuous, integer, and categorical) design variables, which are common in engineering and industrial applications.
The current sampling-based methodology relies on gradient-based SVGD updates and, therefore, assumes that the acquisition function is defined on a continuous design space and is differentiable.
For mixed-type factors, the acquisition function is defined partly on a discrete space, so the usual gradient used in SVGD is no longer available.
To resolve this incompatibility, we may consider mixed-type sampling methods, such as gradient-free Stein variational algorithms \citep{han2018stein,han2020stein} and the mixed Hamiltonian Monte Carlo algorithm \citep{zhou2020mixed}.



\textbf{Data Availability Statement}
The data in this paper are available upon request.
\vspace{-.1in}

\textbf{Declaration of Generative AI Use}
GPT 5.6 Sol was used only for language editing and code-structure reorganization.
\vspace{-.2in}

\renewcommand{\bibfont}{\footnotesize}  
\setlength{\bibsep}{2pt} 
\bibliography{bibliography.bib}

\clearpage
\pagenumbering{arabic}
\appendix
\counterwithin{equation}{section}
\counterwithin{figure}{section}

\begin{center}
    \vspace*{2cm}
    {\LARGE Supplementary Materials for ``Sampling-Based Batch Sequential Design by Stein Variational Gradient Descent''\par}
    \vspace{.2in}
    \if1\anon
    Penghui Fu, Xiaoxian Ding, Chunlin Ji, Jianhua Z. Huang, C. F. Jeff Wu\footnotemark[1]
    \fi
\end{center}
\if1\anon
\footnotetext[1]{Corresponding author: \href{mailto:jeffwu@cuhk.edu.cn}{jeffwu@cuhk.edu.cn}}
\fi

\section{The gradient of the batch ALM criterion}
\label{sec:batch-alm-appendix}
In this section, we derive the gradient of the batch ALM criterion \eqref{eq:hrk-batch-logdet} under the HRK model.
We follow the same notation (\(\mR\), \(\mR_b\), \(\mR_{nb}\), \(\mR_{b\mid n}\), \(\mD_b\)) in \cref{sec:ALM-nonstationary},
and let \(\vr_j\defeq\vr(\vx_{n+j})\) be the \(j\)-th column of \(\mR_{nb}\) and
\(\nabla_j\) be a shorthand for \(\nabla_{\vx_{n+j}}\).
We focus on the noise-free case.
In practice, we include a nugget ratio \(\eta\) for numerical stability, and it suffices to replace \(\mR_{b\mid n}\) by \(\mR_{b\mid n}+\eta\mI_q\).
With the plug-in parameters fixed, \eqref{eq:hrk-batch-logdet} becomes
\[
    \log\det\Cov\!\left(\vf_b\given\calD_n\right)
    =
    2\sum_{j=1}^{q}\log\tau(\vx_{n+j})
    +
    \log\det\mR_{b\mid n}.
\]
For the log-scale term, by \eqref{eq:hrk-model}, we have
\begin{equation}\label{eq:appendix-grad-log-tau}
    \nabla_j\log\tau(\vx_{n+j})
    = -\frac{(\nabla_j\vr_j)\T\vc}{c_0+\vr_j\T\vc} .
\end{equation}
For the log-determinant term, the \((j,k)\)-th entry of \(\mR_{b\mid n}\) is
\begin{equation}\label{eq:appendix-Rbn-entry}
    [\mR_{b\mid n}]_{jk}
    =
    R_{\btheta}(\vx_{n+j},\vx_{n+k})
    -
    \vr_j\T\mR^{-1}\vr_k,
\end{equation}
which implies that only the \(j\)-th row and column of \(\mR_{b\mid n}\) depend on \(\vx_{n+j}\).
Using \(\diff\log\det(\mA)=\operatorname{tr}\left(\mA^{-1}\diff\mA\right)\), we have
\begin{equation}\label{eq:appendix-grad-row-column}
    \nabla_j\log\det(\mR_{b\mid n})
    =
    [\mR_{b\mid n}^{-1}]_{jj}\,\nabla_j[\mR_{b\mid n}]_{jj}
    +
    2\sum_{k=1}^q \bbI\{k\neq j\}\,
        [\mR_{b\mid n}^{-1}]_{jk}\,\nabla_j[\mR_{b\mid n}]_{jk} .
\end{equation}
Differentiating \eqref{eq:appendix-Rbn-entry} with respect to $\vx_{n+j}$ gives, for \(k\neq j\),
\(\nabla_j[\mR_{b\mid n}]_{jk}=\nabla_j R_{\btheta}(\vx_{n+j},\vx_{n+k})-(\nabla_j\vr_j)\T\mR^{-1}\vr_k\),
and for $k=j$, by \(R_{\btheta}(\vx,\vx)=1\),
\(\nabla_j[\mR_{b\mid n}]_{jj}=-2(\nabla_j\vr_j)\T\mR^{-1}\vr_j\). 
Substituting both into
\eqref{eq:appendix-grad-row-column} yields
\begin{equation}\label{eq:appendix-grad-compact}
    \nabla_j\log\det(\mR_{b\mid n})
    =
    2\sum_{k=1}^{q}[\mR_{b\mid n}^{-1}]_{jk}
    \left\{
    \nabla_j R_{\btheta}(\vx_{n+j},\vx_{n+k})
    -
    (\nabla_j\vr_j)\T\mR^{-1}\vr_k
    \right\}.
\end{equation}
Combining \eqref{eq:appendix-grad-log-tau} and \eqref{eq:appendix-grad-compact},
the HRK gradient is
\begin{equation*}
    \nabla_j\log\det\Cov\!\left(\vf_b\given\calD_n\right)
    =
    -2\,\frac{(\nabla_j\vr_j)\T\vc}{c_0+\vr_j\T\vc}
    +
    2\sum_{k=1}^{q}[\mR_{b\mid n}^{-1}]_{jk}
    \left\{
    \nabla_j R_{\btheta}(\vx_{n+j},\vx_{n+k})
    -
    (\nabla_j\vr_j)\T\mR^{-1}\vr_k
    \right\}.
\end{equation*}
The gradient of the batch ALM criterion under the OK model can be obtained by letting $\tau(\vx)\equiv \nu$, which yields
\begin{equation*}
    \begin{split}
    \nabla_j\log\det\Cov\!\left(\vf_b\given\calD_n\right)
    &= \nabla_j\log\det(\mR_{b\mid n}) \\
    &= 2\sum_{k=1}^{q}[\mR_{b\mid n}^{-1}]_{jk}
    \left\{
    \nabla_j R_{\btheta}(\vx_{n+j},\vx_{n+k})
    - (\nabla_j\vr_j)\T\mR^{-1}\vr_k
    \right\}.
    \end{split}
\end{equation*}

\section{Information-based criteria and batch extensions}\label{sec:info-criteria}
In this section, we review the information-based criteria for fully sequential designs and their batch extensions by definition.
As mentioned in Section~\ref{sec:extension}, information-based criteria select inputs to maximize information about quantities of interest $\btheta$, which can be measured in either a Bayesian or Frequentist manner.

Bayesian adaptive design selects the next point $\vx_{n+1}$ to maximize the expected information gain \citepsupp[EIG,][]{lindley1956measure}, which is defined as the reduction of entropy of $\btheta$, after observing $y_{n+1}$:
\begin{equation}\label{eq:seq-EIG}
    \begin{split}
        \text{EIG}(\vx_{n+1}\mid\calD_n) &:= H(p(\btheta\mid\calD_n)) - \bbE_{y_{n+1}}[H(p(\btheta\mid\calD_n\cup \{\vx_{n+1},y_{n+1}\}))] \\
        &= \bbE_{\btheta,y_{n+1}}\!\left[\log p(y_{n+1}\mid\btheta,\vx_{n+1},\calD_n)-\log p(y_{n+1}\mid\vx_{n+1},\calD_n)\right].
    \end{split}
\end{equation}
Here, $H$ denotes the entropy, and $\bbE_{\btheta,y_{n+1}}$ is the expectation taken with respect to $p(\btheta, y_{n+1}\mid\calD_n, \vx_{n+1})$.
The EIG (\ref{eq:seq-EIG}) can be directly extended to a batch version in a similar manner as ALC \citepsupp{kirsch2019batchbald}
\begin{equation*}
    \begin{split}
    \text{Joint-EIG}(\mX_b\mid\calD_n) &:= H(p(\btheta\mid\calD_n)) - \bbE_{\vy_b}[H(p(\btheta\mid\calD_n\cup \{\mX_b,\vy_b\}))] \\
    &= \bbE_{\btheta,\vy_b}\!\left[\log p(\vy_b\mid\btheta,\mX_b,\calD_n)-\log p(\vy_b\mid\mX_b,\calD_n)\right],
    \end{split}
\end{equation*}
where $\bbE_{\btheta,\vy_b}$ is the expectation taken with respect to $p(\btheta, \vy_b\mid\mX_b, \calD_n)$.


Alternatively, the sequential Fisher information (FI)-based design selects the next point to maximize the Fisher information of $\btheta$ from the new data $\{\vx_{n+1},y_{n+1}\}$ .
The idea can be extended to batch sequential designs:
\begin{equation}\label{eq:batch-FI}
    \begin{split}
        \mX_b^\star = \argmax_{\mX_b}\, \mathcal{F}(\text{FIM}(\hat\btheta_{\calD_n}, \mX_b)) = \argmax_{\mX_b}\,\mathcal{F}\left[\left.-\bbE_{\vy_b}\left(\frac{\partial^2 \log p(\vy_b|\mX_b,\calD_n,\btheta)}{\partial \btheta^2}\right|_{\hat\btheta_{\calD_n}}\right) \right].
    \end{split}
\end{equation}
Here, $\text{FIM}$ denotes the Fisher information matrix; 
$\mathcal{F}$ is a summary statistic;
$\bbE_{\vy_b}$ is the expectation with respect to $p(\vy_b|\mX_b,\calD_n,\hat\btheta_{\calD_n})$; 
$\hat\btheta_{\calD_n}$ is a point estimate of $\btheta$ based on $\calD_n$.
Intuitively, (\ref{eq:batch-FI}) selects $\mX_b$ such that $\vy_b$ leads to the most accurate estimate of $\btheta$. 

\section{Space-filling criteria and batch extensions}\label{sec:space-filling-criteria}
In this section, we review the minimum-energy design \citepsupp[MED,][]{joseph2015sequential} and the maximum-projection design \citepsupp[MaxPro,][]{joseph2015maxpro}, and discuss their batch extensions by definition.
MED views points as charged particles in a box and minimizes the total electric potential energy, which is defined as the summation of potential energy $\text{PE}(\vx,\vx') = (\rho(\vx)\rho(\vx')/\norm{\vx-\vx'})^{k}$ for all pairs of different points $\vx$ and $\vx'$ in the box.
Here, $\rho(\vx)>0$ is the charge function that models the relative importance of point $\vx$; $k>0$ is a pre-specified power index.
Given the current data $\calD_n$, the sequential MED selects the next point by minimizing the increased total energy after augmenting $\calD_n$ with $\vx_{n+1}$:
\begin{equation}\label{eq:seq-MED}
    \argmin_{\vx_{n+1}}\,\sum_{1\leq i<j\leq n+1} \text{PE}(\vx_i,\vx_j) - \sum_{1\leq i<j\leq n} \text{PE}(\vx_i,\vx_j) = \argmin_{\vx_{n+1}}\, \sum_{i=1}^{n} \text{PE}(\vx_i,\vx_{n+1}).
\end{equation}
Following the same idea, batch MED \citepsupp{kim2017batch} can be defined to minimize the increased total energy after including a batch of $q$ points, i.e.,
\begin{equation}\label{eq:batch-MED}
    \begin{split}
         &\argmin_{\mX_b}\,\sum_{1\leq i<j\leq n+q} \text{PE}(\vx_i,\vx_j) - \sum_{1\leq i<j\leq n} \text{PE}(\vx_i,\vx_j) \\
         &= \argmin_{\mX_b}\, \sum_{j=n+1}^{n+q}\sum_{i=1}^{n} \text{PE}(\vx_i,\vx_{j}) + \sum_{n+1\le i < j \le n+q} \text{PE}(\vx_{i},\vx_{j}).
    \end{split}
\end{equation}
Regarding MaxPro designs, the one-shot criterion for $n$ design points $\{\vx_i\}_{i=1}^n$ is
\begin{equation}\label{eq:maxpro}
    \sum_{1\leq i<j\leq n} \frac{1}{\prod_{l=1}^{d}(x_{il}-x_{jl})^2}.
\end{equation}
Here, $x_{il}$ and $x_{jl}$ are the $l$-th entries of $\vx_i$ and $\vx_j$, respectively. 
Since the one-shot criterion (\ref{eq:maxpro}) takes a similar additive form to the total potential energy in MED, its fully sequential and batch sequential versions can be defined analogously. We omit the details.

\section{Computational time}
\begin{figure}[!htb]
    \centering
    \begin{subfigure}{0.28\textwidth}
        \includegraphics[width=\textwidth]{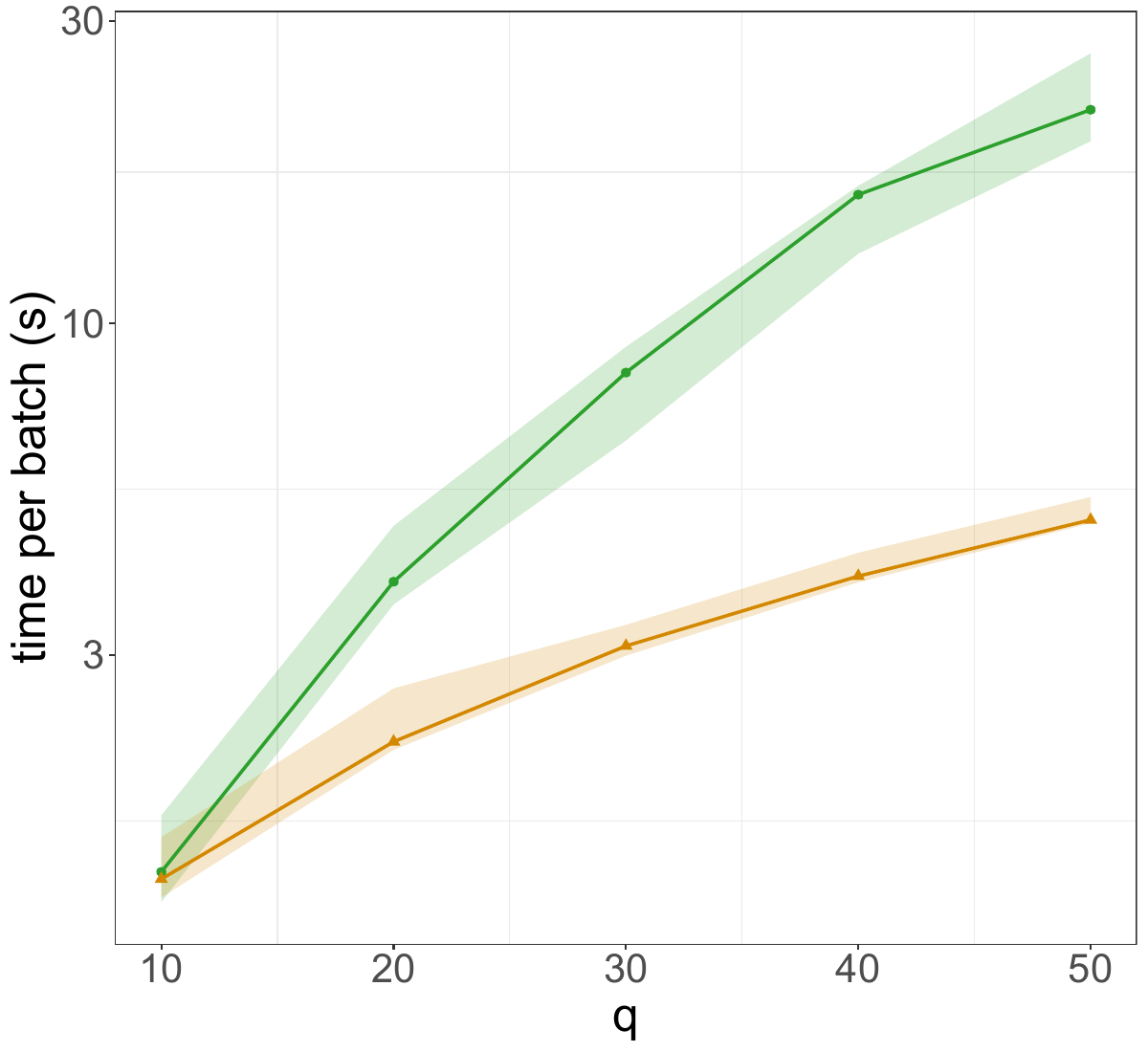}
        \caption{Gramacy--Lee}
    \end{subfigure}\hfill
    \begin{subfigure}{0.28\textwidth}
        \includegraphics[width=\textwidth]{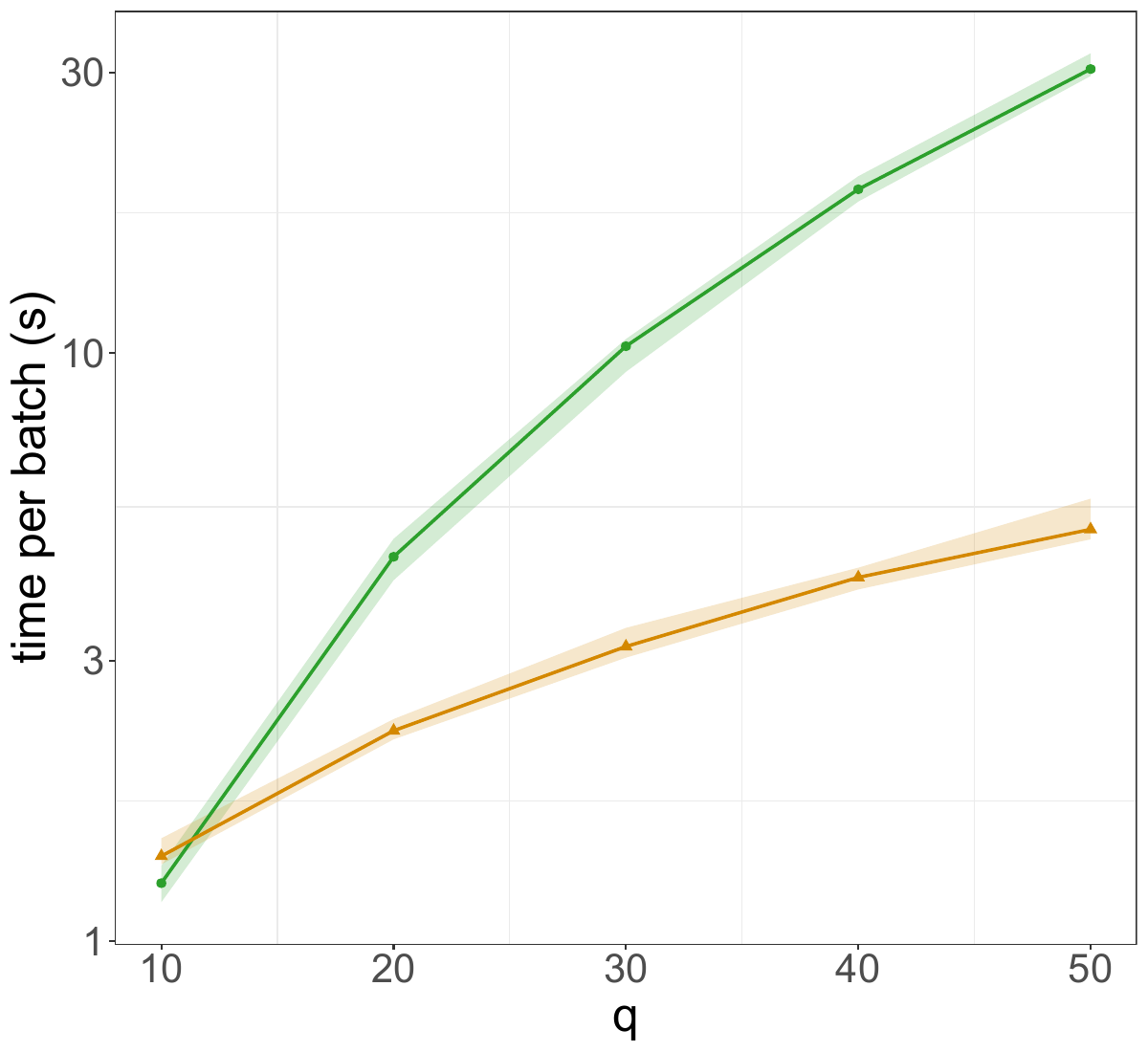}
        \caption{Dette--Pepelyshev}
    \end{subfigure}\hfill
    \begin{subfigure}{0.28\textwidth}
        \includegraphics[width=\textwidth]{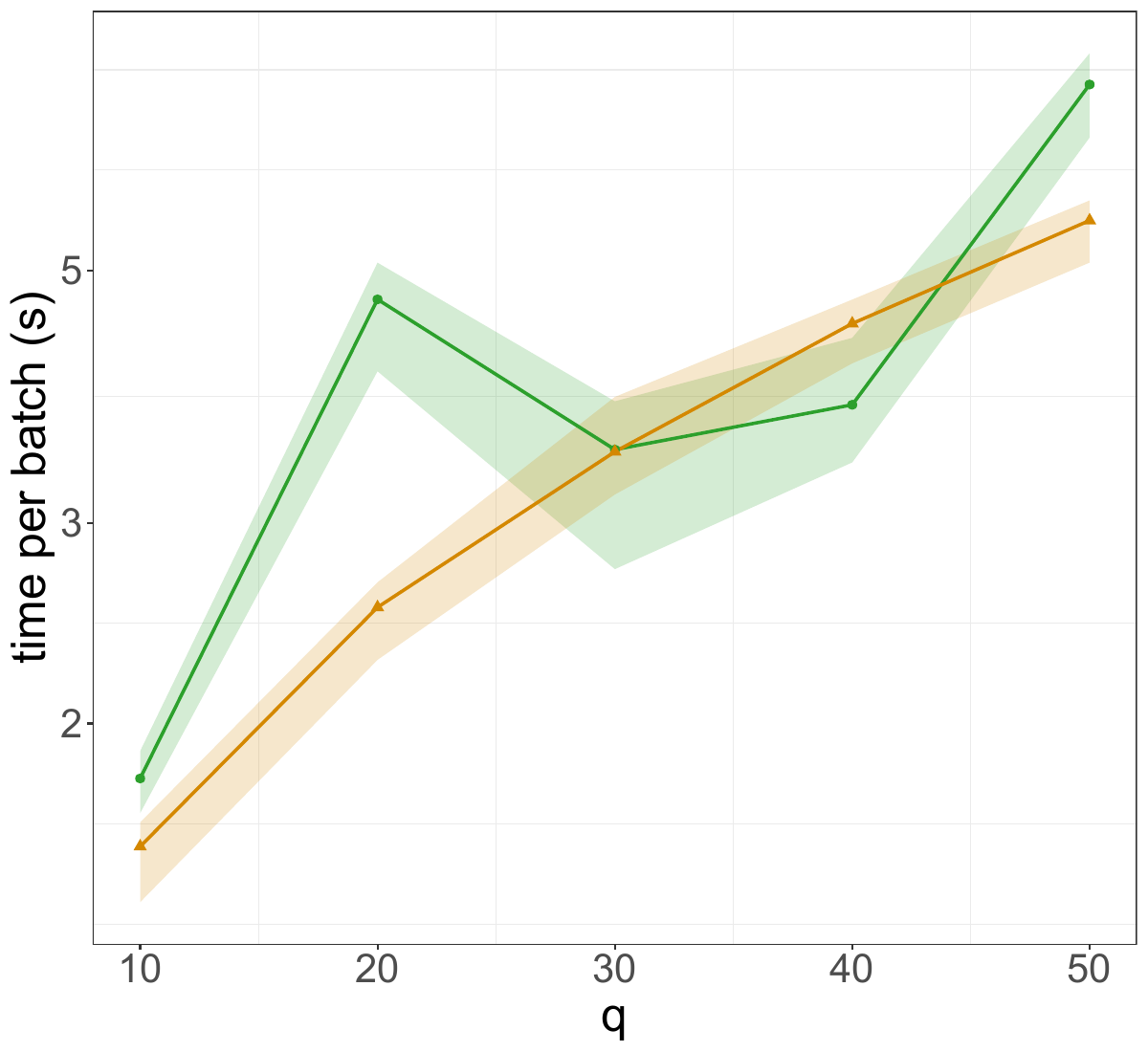}
        \caption{Cantilever beam}
    \end{subfigure}\\[1.5ex]
    \begin{subfigure}{0.28\textwidth}
        \includegraphics[width=\textwidth]{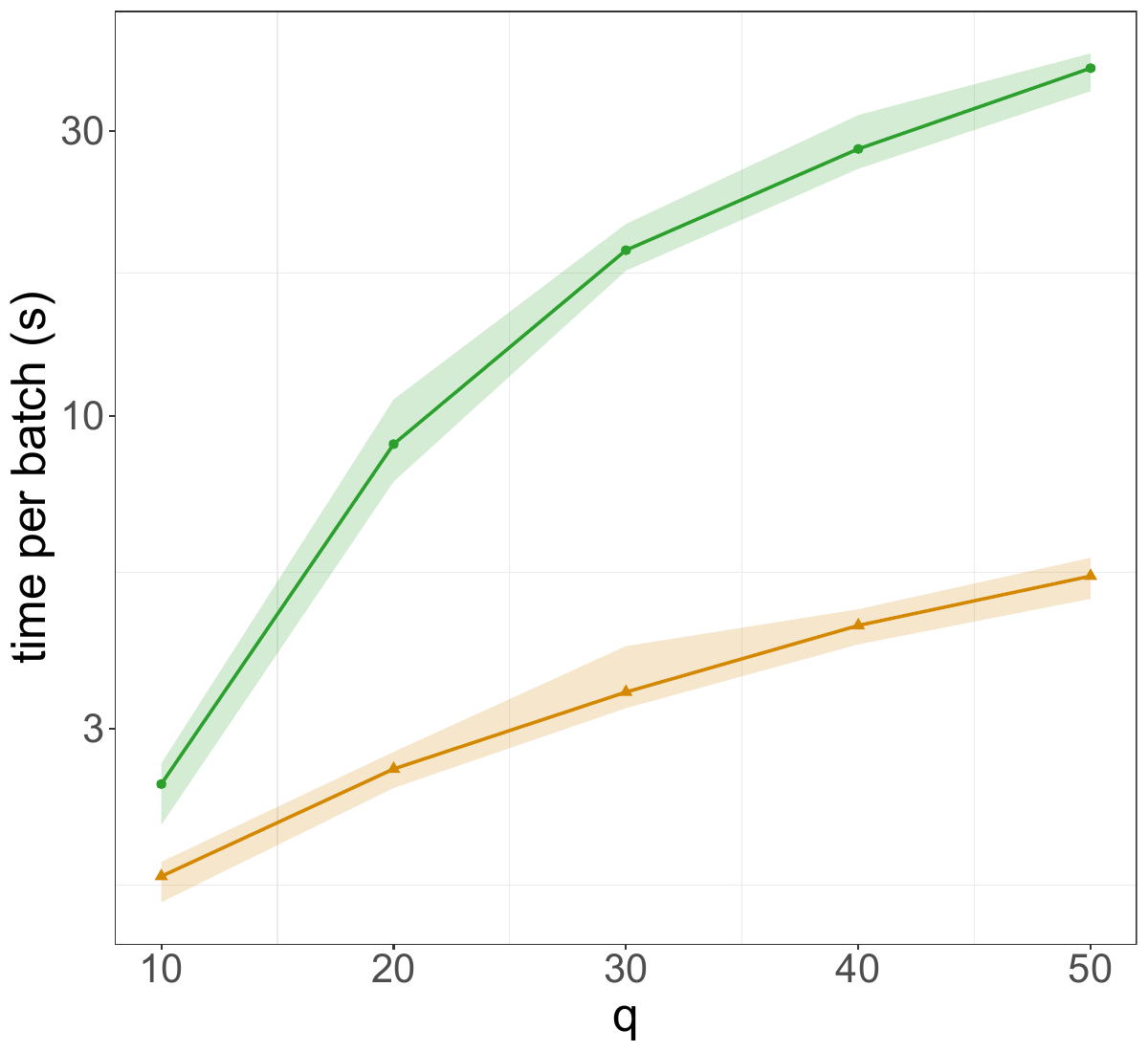}
        \caption{OTL circuit}
    \end{subfigure}\hfill
    \begin{subfigure}{0.28\textwidth}
        \includegraphics[width=\textwidth]{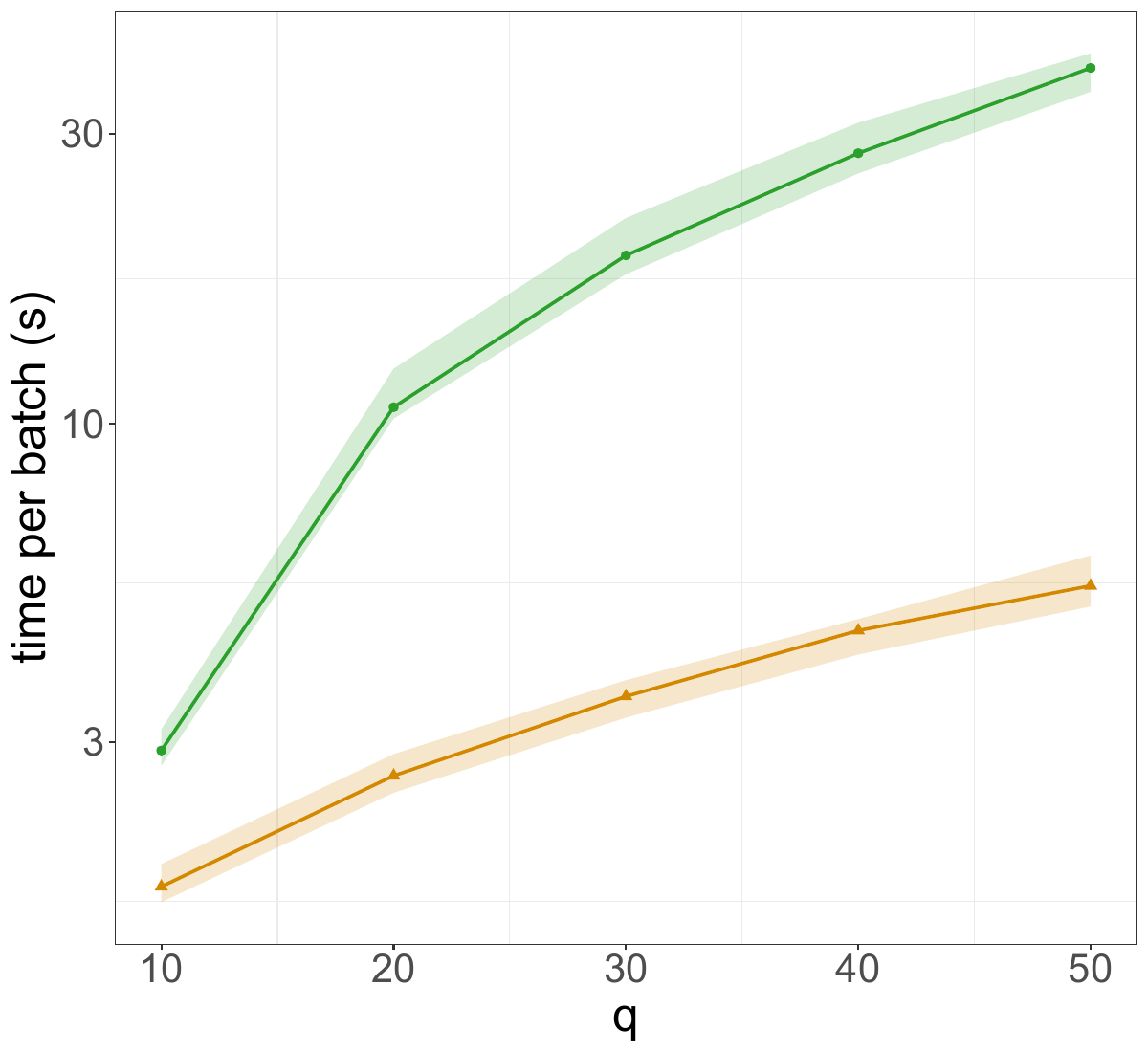}
        \caption{Piston}
    \end{subfigure}\hfill
    \begin{subfigure}{0.28\textwidth}
        \includegraphics[width=\textwidth]{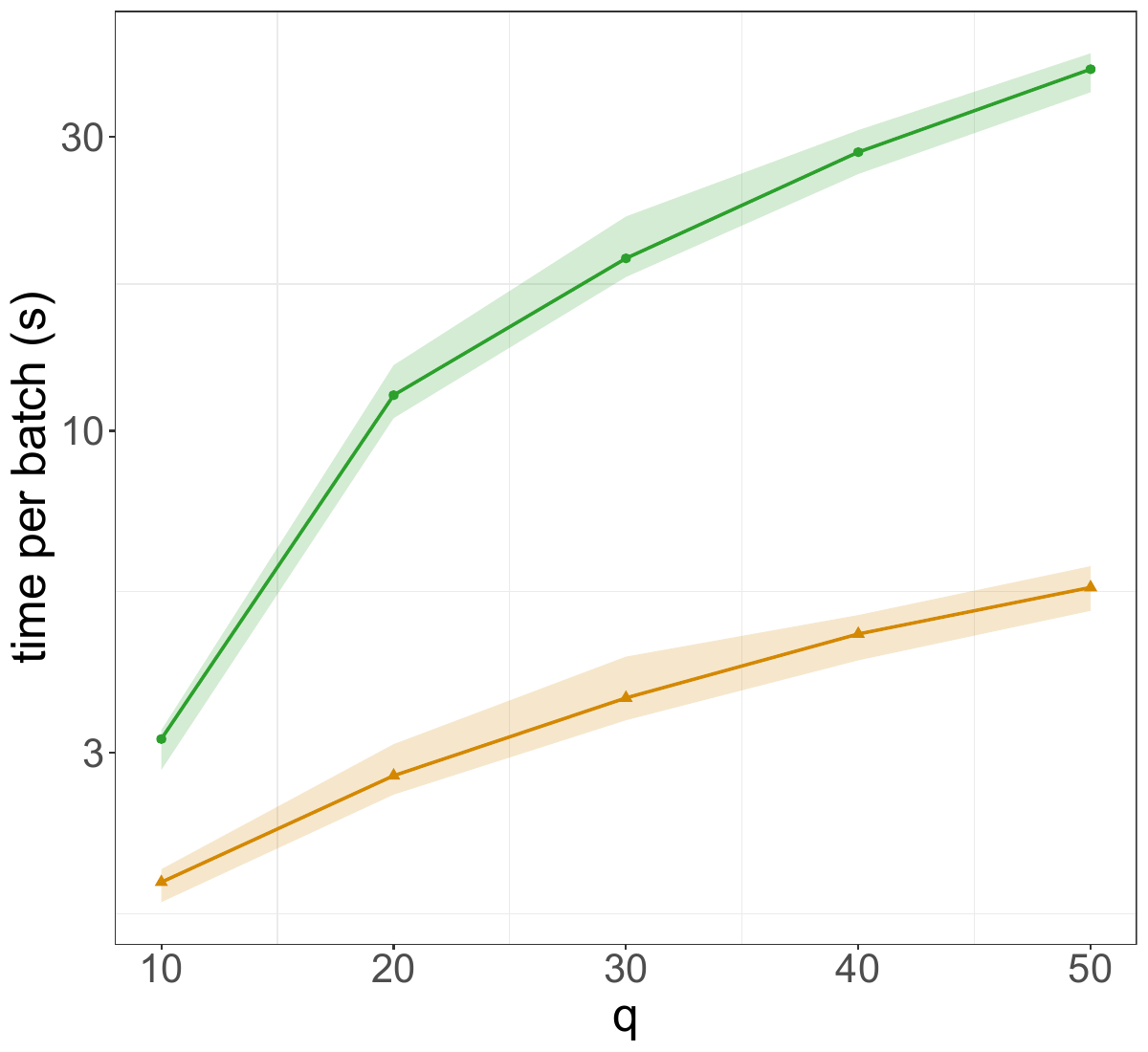}
        \caption{Borehole}
    \end{subfigure}\\[2ex]
    \includegraphics[width=0.55\textwidth]{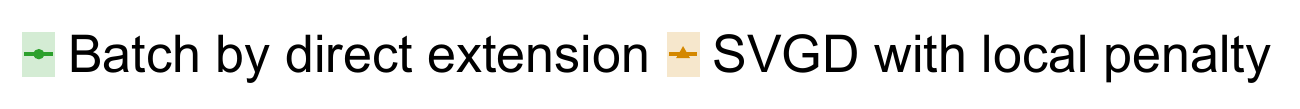}
    \caption{Processing time (log scale) of each batch for two methods. Solid lines represent the median over $10$ repetitions, and the shaded bands mark the $10$th and $90$th quantiles.}
    \label{fig:hrk-time-benchmark}
\end{figure}

Following Section~\ref{sec:ALM-nonstationary}, we compare the computational time for selecting one batch of varying sizes from the same fixed $20$-point MaxPro design.
Other settings remain the same as in Section~\ref{sec:ALM-nonstationary}.
The experiment is repeated for $10$ times due to the randomness in generating the initial MaxPro design.
Figure~\ref{fig:hrk-time-benchmark} shows that batch by SVGD is overall substantially faster than batch by direct extension, especially as the batch size increases.

\bibliographystylesupp{apalike}
\bibliographysupp{bibliography}  
\end{document}